%% file: pca.tex
\documentclass[usenatbib]{mn2e}
\usepackage{amsmath}
\usepackage{natbib}
\usepackage{epsfig}
\usepackage{txfonts}
\usepackage{pict2e}
\usepackage{cleveref}
\usepackage{color}
\newcommand{\dxe}{\Delta \xe /\xe }
\newcommand{\dlnxe}{\Delta\ln\xe}
\newcommand{\ddxe}{\frac{\Delta \xe}{\xe}}

\usepackage{url}
\usepackage{cleveref}
\usepackage{xspace}

\newcommand{\test}[1]{{\textcolor{black}{#1}}}

\newcommand{\aEM}{\alpha_{\rm EM}/\alpha_{\rm EM,0}}
\newcommand{\Atwo}{A_{\rm 2s1s}}
\newcommand{\yp}{Y_{\rm p}}
\newcommand{\fann}{f_{\rm ann}}

\newcommand{\planck}{{\it Planck}\xspace }

\newcommand{\DelL}{\Delta\ln\mathcal{L}}
\newcommand{\omb}{\Omega_{\rm b} h^2}
\newcommand{\omc}{\Omega_{\rm c} h^2}
\newcommand{\thetaMC}{100\,\theta_{\rm MC}}
\newcommand{\LCDM}{$\Lambda$CDM\xspace}
\newcommand{\ns}{n_{\rm s}}
\newcommand{\As}{A_{\rm s}}

\newcommand{\logA}{\ln(10^{10} A_{\rm s})}
\newcommand{\Lvec}[2]{L^{#1}_{#2}}
\newcommand{\parA}{\mathcal{A}}

\newcommand{\delA}{\Delta\parA}
\newcommand{\ddA}{\frac{\delA}{\parA}}
\newcommand{\Corr}[1]{{\rm Corr\left(#1\right)}}

\input{Befehle}
\usepackage[none]{hyphenat} 
\renewcommand{\xe}{X_{\rm e}} 

\title[Improved recombination constraints]
{Improved model-independent constraints on the recombination era and development of a direct projection method}

\author[L.~Hart and J.~Chluba]{
Luke Hart$^{1}$\thanks{Email: luke.hart@manchester.ac.uk} and Jens Chluba$^{1}$ 
\\
$^{1}$Jodrell Bank Centre for Astrophysics, Alan Turing Building, University of Manchester, Manchester M13 9PL \\}

\date{\vspace{-5mm}Accepted  --. Received --.}

\pubyear{2019}

\begin{document}
\label{firstpage}
\pagerange{\pageref{firstpage}--\pageref{lastpage}}
\maketitle
\begin{abstract}
The unparalleled precision of recent experiments such as \planck have allowed us to constrain standard and non-standard physics (e.g., due to dark matter annihilation or varying fundamental constants) during the recombination epoch. 
However, we can also probe this era of cosmic history using model-independent variations of the free electron fraction, $\xe$, which in turn affects the temperature and polarization anisotropies of the cosmic microwave background. In this paper, we improve on the previous efforts to construct and constrain these generalised perturbations in the ionization history, deriving new optimized eigenmodes based on the full \planck~2015 likelihood data, introducing the new module {\tt FEARec++}. We develop a direct likelihood sampling method for attaining the numerical derivatives of the standard and non-standard parameters, and discuss complications arising from the stability of the likelihood code. We improve the amplitude constraints of the \planck 2015 principal components constructed here, $\mu_1=-0.09\pm0.12$, $\mu_2=-0.17\pm0.20$ and $\mu_3=-0.30\pm0.35$, finding no indication for departures from the standard recombination scenario. The error constraint on the third mode has been improved by a factor of $2.5$. We utilise an efficient eigen-analyser that keeps the cross-correlations of the first three eigenmodes to $\Corr{\mu\,\mu'}<0.1\%$ after marginalisation for all the considered data combinations. We also propose a new projection method for estimating constraints on the parameters of non-standard recombination scenarios.
As an example, using our eigenmode measurements this allows us to recreate the \planck constraint on the two-photon decay rate, $\Atwo=7.60\pm0.64$, giving an error estimate to within $\simeq 0.05\sigma$ of the full MCMC result. The improvements on the eigenmodes analysis using the \planck data will allow us to implement this new method for analysis with fundamental constant variations in the future.
\end{abstract}

\maketitle
\section{introduction}
   Efforts to measure the anisotropies of the cosmic background radiation (CMB) in recent years have allowed us to constrain parameters describing the matter content, primordial fluctuations and reionization epoch of the Universe with exceptional precision \citep{wmap9params, Planck2015params, Planck2018params}.  Since the data is now sensitive to sub-percent level effects in the recombination dynamics \citep{Fendt2009, Jose2010, Shaw2011}, there has been a need for advanced recombination codes such as {\tt CosmoRec} \citep{Chluba2010b} and {\tt HyRec} \citep{Yacine2010c}. These carefully capture the  atomic physics and radiative transfer processes in the recombination epoch. 
   For example, the induced two-photon decay process and time-dependence of electronic level populations lead to corrections at the $1\%$ level to the free electron fraction \citep{Chluba2006, Chluba2007, Grin2009, Yacine2010}. 
 The sensitivity to small modifications to the recombination dynamics around redshift $z\simeq 1100$ has even lead to a measurement of the hydrogen 2s-1s two-photon decay rate using \planck data \citep{Planck2015params}, confirming the theoretical value at $\lesssim 7\%$ precision.
      
While measurements of the CMB have given us great insights into the standard $\Lambda$CDM cosmological framework, they have also allowed us to constrain extensions to the standard paradigm. Constraints on neutrino species and their masses \citep[][for more details]{PlanckNeutrino, BattyeNeutrinos, Abazajian2015} have been studied in great detail and  tests of Big Bang Nucleosynthesis have also been considered \citep{Coc2013, PRISM2013WPII, CMBS42016}. Some non-standard physical extensions have unique effects on the ionization history, for instance, allowing us to investigate the effects of {\it dark matter annihilation / decay} \citep{Chen2004, Slatyer2009, Finkbeiner2012, Galli2013, Slatyer2016}, {\it variations of fundamental constants} \citep{Battye2001, Scoccola2009, Planck2015var_alp, Hart2017} and {\it primordial magnetic fields} \citep[e.g.,][]{Sethi2005, Shaw2010PMF, Kunze2014, Chluba2015PMF, Planck2016PMF}.
   
All the above examples are based on parametric extensions of the standard cosmological model. However, in particular for effects directly stemming from variations of the cosmological recombination history, an alternative approach is possible.
Small variations of the free electron fraction, $\xe$, lead to changes of the Thomson visiblity function and this directly affects the CMB temperature and polarization anisotropies. Individual changes in narrow redshift ranges usually yield far too low a signal to be constrained; however, by analysing the correlated responses across many redshifts, one can find optimal $\xe$ eigenmodes. 
This process is called a {\it principal component analysis} (PCA) and has been extensively used in cosmology \citep[see][for various physically-motivated examples using PCA]{Mortonson2008, Finkbeiner2012, Dai2018, Campeti2019}.
By creating an eigenbasis specifically focused on variations to the ionization history in redshift space, we can use CMB data to constrain the corresponding principal component amplitudes. This idea has been covered previously \citep{Farhang2011,Farhang2013}, and direct constraints using \planck CMB data were derived \citep{Planck2015params, Planck2018params}, showing that no significant departures from the standard recombination scenario are expected. 
   
However, a couple of aspects deserve further investigation. While the results from previous works performed as expected for idealized experiments, some of the finer computational points such as the time-step during recombination and its effect on the numerical stability of basis functions and eigenmodes were not elaborated on. Similarly, optimisation of the modes explicitly using the likelihood code were not carried out in as much detail. 
These in turn lead to spurious correlations between eigenmodes and larger correlations for the standard parameters, which also increased the marginalized errors of the modes. This motivated us to reassess the situation to improve an extend previous recombination analysis.  
 
In the present paper, we explain the formalism of a new analytical and numerical Fisher matrix method for the creation of ionization history eigenmodes. This introduces the recombination module {\tt FEARec++} (Fisher Eigen Analyser for Recombination), with associated modifications to {\tt CAMB} \citep{CAMB}, {\tt CosmoMC} \citep{COSMOMC} and {\tt CosmoRec} \citep{Chluba2010}.
{\tt FEARec++} can be used to attain both cosmic-variance limited (CVL) and direct-likelihood sampled eigenmodes, overcoming some of the aforementioned limitations (Sec.~\ref{sec:formalism}). 
The recombination eigenmodes for both CVL and likelihood sampling are presented in Sec.~\ref{sec:recomb} along with comparisons to analysis done in previous works \citep[i.e.,][]{Farhang2011, Planck2015params}. 
We then present the MCMC constraints and contours for these converged eigenmode amplitudes along with a discussion of the residual mode correlations (Sec.~\ref{sec:mcmc}). 

Finally, we introduce a novel projection method that uses the obtained mode constraints to quickly estimate errors for non-standard recombination scenarios in Sect.~\ref{sec:proj}. The strongly constrained, orthogonal, non-parametric principal components can be used with a given parametrised physical variation in the same quantity to attain convincing constraints for the associated parameters, without the requirement for MCMC on a case-by-case basis. The limitations to this direct projection method are outlined along with our wider plan of incorporating this analysis into a follow-up paper for varying fundamental constants. 
Additional extensions and updates regarding the \planck 2018 data, which became available this summer, will be left for future work.

\vspace{-4mm}
\section{Formalism}
\label{sec:formalism}
The formalism of the PCA depends on variations in an observable quantity as a function of a given parameter. These changes are compared to the expected errors in the measurement. Here, our observables are the CMB temperature and polarisation power spectra, quantified by $C_l^{\rm X}$, where $X = \{TT,TE,EE\}$, and we are varying the relative value of the free electron fraction, $\xe(z)$, at various redshifts $z$.

Firstly, we create a generalised function space over a given redshift range $z$. We fill this redshift range with a full set of basis functions $\varphi_i(z)$ defined as,
\begin{equation}
    \varphi_i(z) = \frac{\Delta\xe}{\xe}(z,z_i) \simeq \dlnxe(z,z_i).
\label{eq:basis}
\end{equation}
Each basis function is associated with a given pivot redshift, $z_i$ and shape. Previous works have shown that different functional forms (i.e. Gaussians, M4 splines, Chebyshev polynomials) lead to the same modes, assuming that they are orthogonal \citep[see][for more details]{Farhang2011}. For our analysis, we shall use a set of Gaussian basis functions, as described in Appendix~\ref{app:basis}. There we also explain how the set of Gaussians can be used to construct the orthonormal set of principle components in $\dlnxe(z)$.
The perturbations, $\varphi_i(z)$, are then added to the recombination output of {\tt CosmoRec}. Here, in particular we use the submodule {\tt Recfast++}. It has already been demonstrated how the use of a correction function with {\tt Recfast++} gives similar results to {\tt CosmoRec} for small deviations from the standard model \citep{Jose2010, Shaw2011, Hart2017}.

\begin{figure}
  \centering
  \includegraphics[width=0.95\linewidth, trim={0 28 0 0},clip]{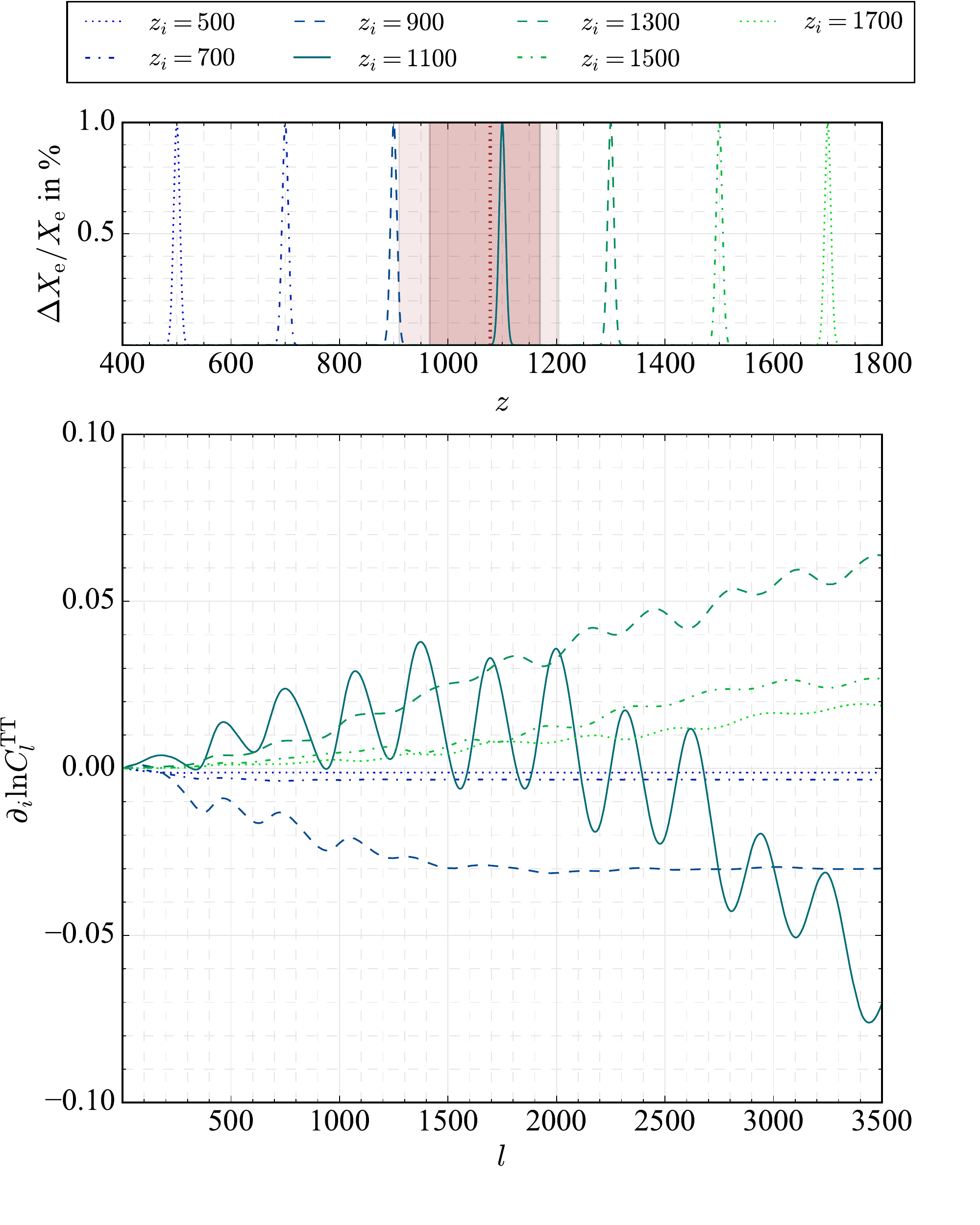}
  \caption{Relative differences in the CMB temperature power spectrum from Gaussian basis functions pivoted around various redshifts in the range $z_i \in [500,1700]$. For context, the redshift of the maximum of the Thomson visibility function for a standard $\Lambda$CDM cosmology has been included with the half maximum width and quarter maximum width of the function .}
  \label{fig:responses}
\end{figure}

\vspace{-0mm}
\subsection{Analytical approach}
\label{sec:analytic_method}
In this section, we define the analytical method for generating principal components of $\dxe$ across the recombination era. This is particularly useful for computations of cosmic-variance limited modes, but some of the modifications to the Boltzmann solver are also crucial when directly working with the likelihood function as explained in the next section.

The changes in the ionization history, defined by Eq.~\eqref{eq:basis}, are propagated through to the CMB anisotropies via the Thomson visibility function with the Boltzmann code {\tt CAMB} \citep{CAMB}. For every given basis function, $\varphi_i(z)$, we can measure the relative difference in the CMB power spectrum, $\partial C_l/\partial p_i$, where $p_i$ represents the amplitude of the basis function which perturbs the ionization history.

Numerical derivatives of the CMB temperature power spectrum from these basis functions are shown in Fig.~\ref{fig:responses}. 
Their precise computation required several modifications to {\tt CAMB} as explained in Appendix~\ref{app:analytic}. In particular work on the time-sampling across the recombination era was found to be crucial for obtaining numerically-stable responses.
The responses for the CMB spectra are shown as derivatives such that $\partial_i\ln C_l^{TT} = \partial C_l^{TT}/\partial p_i \, (1/C_l^{ TT,0})$, where $C_l^{ TT,0}$ refers to the fiducial CMB power spectrum given a typical set of standard $\Lambda$CDM parameters. The strongest responses in the CMB temperature power spectrum occur near the \test{peak} of the Thomson visibility function ($z_{\rm rec}\simeq 1100$), with a positive change in the slope of the spectra for $z>z_{\rm rec}$ and a negative change in the slope for $z<z_{\rm rec}$. This behaviour is reflected in the changes found from the polarisation spectra as well.\footnote{Short movie of the responses across the full redshift range can be found at\; \url{ https://sites.google.com/view/pca-recombination/}}

From the CMB responses for given basis functions, we next generate a Fisher matrix, $F_{ij}$, using an analytical approach that is described in Appendix~\ref{app:analytic}. To constrain the most-responsive changes we can diagonalise the matrix to generate a matrix of eigenvalues, $f_{ij} = {\rm diag}\left(\lambda_i\right)$, and a matrix of eigenvectors $M_{ik}$ such that,
\begin{equation}
    F_{ij} = M_{ik}\,f_{kl}\,M^{\rm T}_{lj}.
\end{equation}
We can now recast the basis functions onto the eigenvectors defining $E_k(z)$, 
\begin{equation}
  E_k(z) = \sum_{i=1}^N M_{ik}\varphi_i(z).
\end{equation}
The set of eigenvectors $E_k(z)$ reconstructed from the basis functions are called {\it principal components}. This gives us a set of vectors in $\xe(z)$ space; however, we need smoothed, continuous functions of $\xe$ to effectively solve the Boltzmann equations with the routines supplied in codes such as {\tt CAMB}. For this, we implement an interpolation scheme which preserves orthogonality of the original basis, generates continuous eigenmodes and retains the correct amplitudes of each of the basis functions. After the interpolation, all the principal components are normalised such that,
\begin{equation}
\int_{z_{\rm min}}^{z_{\rm max}}E^2_{\,k}(z){\rm d}z=1.
\end{equation}
where $z_{\rm min}$ and $z_{\rm max}$ define the range of our redshift space. This has been done using Simpson's rule for numerical integration leading to normalisation with an error $\simeq 0.008\%$. The whole procedure is explained in more detail in Appendix~\ref{app:basis}.

\vspace{-0mm}
\subsection{Likelihood sampling approach}
\label{sec:direct_method}
While the direct analytical method is useful for creating cosmic variance limited modes, for applications to a given dataset it is better to directly use the associated likelihood code. For our analysis of \planck, we thus also directly sourced the likelihood function to generate principal components using the formal definition of the Fisher matrix,
\begin{equation}
    F_{ij} = \left<\frac{\partial^2\ln\mathcal{L}}{\partial p_i \partial p_j}\right>_{\mathcal{L} = \mathcal{L}_0},
    \label{eq:fij_definition}
\end{equation}
where $\ln\mathcal{L}$ is the log-likelihood function we are sourcing. \test{Note that the deviations of the likelihood are taken around the fiducial best-fit cosmology represented by $\mathcal{L}_0$}. In contrast to the analytical method, we no longer look for the differential responses in the CMB power spectrum, but we use the likelihood values for a given dataset and group of model parameters. To circumvent the need for re-evaluating foregrounds, instrumental noise and other systematics, we can directly call the likelihood function from an MCMC code [i.e. {\tt CosmoMC} or {\tt MontePython} \citep{Brinckmann2018}]. This requires a careful \test{\emph{stability analysis}} of the numerical derivatives and the precision of the likelihood code.  However this method allows the marginalisation over the standard cosmological and nuisance parameters for any combination of datasets. This is quite powerful since it yields eigenmodes that are decorrelated from standard parameter variations, foreground variations and other correlations that the likelihood is reliant upon.  All the finer details of this method are explained in Appendix \ref{app:direct}. In particular the numerical accuracy of the \planck likelihood function imposed a challenge. 
Together with interpolation of the obtained eigenvectors we then obtain optimal recombination PCs for various data combination.

\begin{figure}
    \centering
    \includegraphics[width=0.92\linewidth]{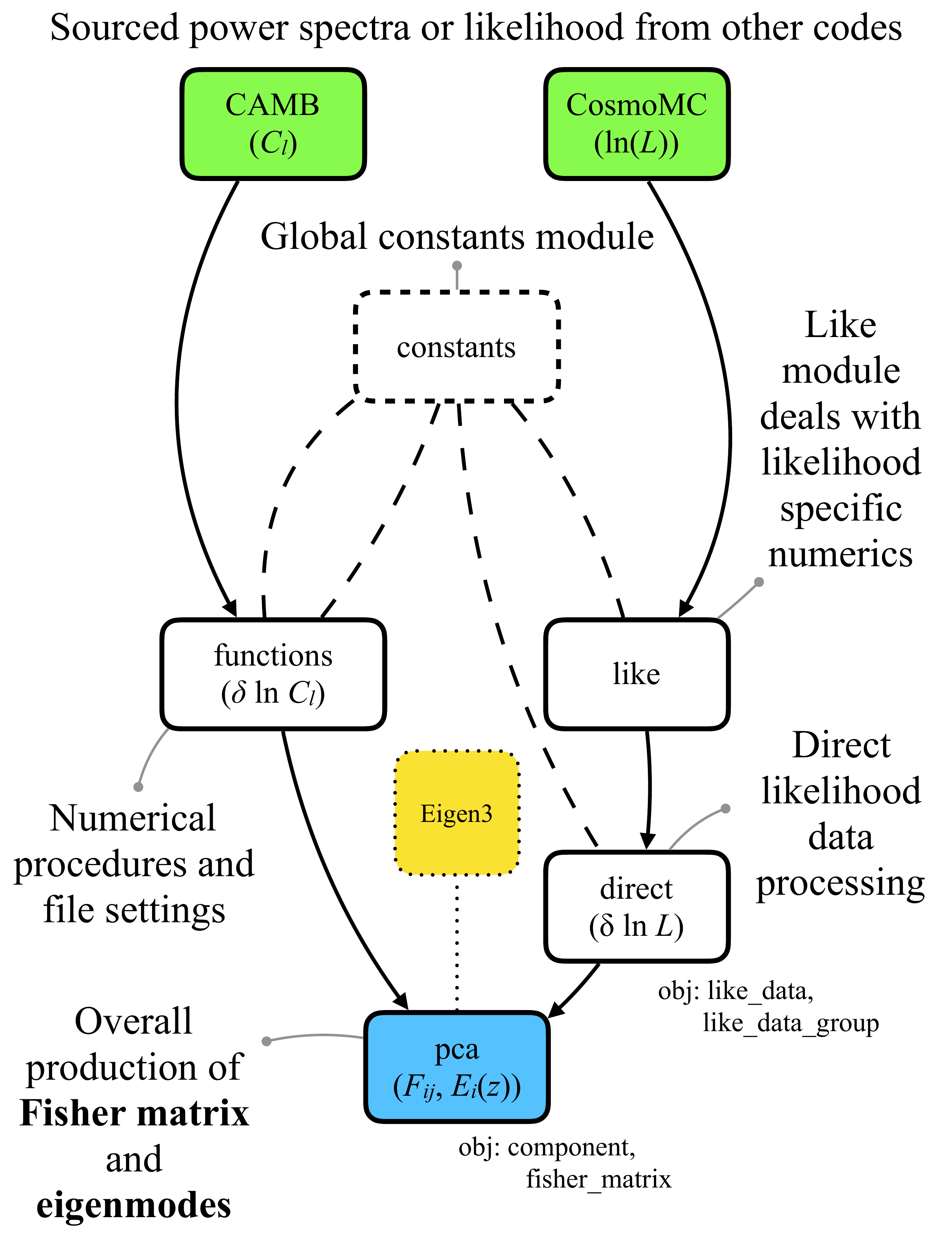}
    \caption{Flow chart of the code {\tt FEARec++} designed to compute the Fisher matrix and subsequent eigenmodes quickly and efficiently. This code has capabilities for analytic Fisher matrices and numerical methods directly using provided likelihoods. {\it Obj} indicates the class objects designed within the appropriate modules. The inclusion of {\tt Eigen3} has been highlighted here as an external module to the PCA code.}
    \label{fig:flow}
\end{figure}

\vspace{-3mm}
\subsection{FEARec++: Fisher Eigen Analyser for Recombination} 
\label{sec:fearec}
The procedures described above are incorporated by the recombination module {\tt FEARec++}, developed as part of this work. This code has been equipped to deal with generalised changes to the recombination history that allow for analytical tests and numerical analyses with various likelihood combinations. The PCA module was added to {\tt CosmoRec} along with a new module in {\tt CosmoMC}.

For the data analysis, the decorrelated eigenmodes are then added back to the Boltzmann code with a given amplitude $\mu_i$ as,
\begin{equation}
    \xe'(z) =\xe(z)\left(1+ \sum_{i=1}^{N_\mu}\mu_i E_i(z)\right).
\end{equation}
Here $N_\mu$ represents the number of modes included in the analysis. A complete representation if the perturbation is achieved for ${N_\mu\rightarrow\infty}$ \citep[e.g.,][]{Farhang2011}; however, in real applications we restrict ourselves to $N_\mu\leq 3$, since higher order modes become less constrained. As the amplitude $\mu_i$ varies the impact of the principal components, we perform an MCMC analysis to find the best fit values for these parameters. This reveals how model-independent changes fit to the data and whether a non-standard change to recombination is favoured by a given dataset.

A detailed flow chart  of {\tt FEARec++} is displayed in Fig.~\ref{fig:flow}. It demonstrates how the code takes inputs from either a Boltzmann code or likelihood sampler (here the examples are {\tt CAMB} and {\tt CosmoMC}). The results are then numerically modified and stored for the PCA code to use with the {\tt Eigen3} module. This is an  efficient C++ eigen-solver that exploits object oriented programming features \citep{EigenCode}. The Fisher matrix and eigenmodes are then obtained with their predicted eigenvalues. There are optional features to allow for unmarginalised and marginalised outputs, which is useful for comparison.

The {\it selective sampling} module added to {\tt CosmoMC} allows the user to generate the likelihood samples for the derivatives required, assuming that the Boltzmann solver in {\tt CAMB} has been modified accordingly (Appendix~\ref{app:analytic}). This is straightforward for $\Lambda$CDM derivatives and includes a runmode that allows testing of the stability of the derivatives, for the user to confirm the achieved accuracy. The recombination code {\tt CosmoRec} has been altered to allow for general variations in a given parameter. The only requirement from the user, is to apply these functions to their chosen constraining variable and to optionally add this as an alternative run-mode. The altered files of {\tt CAMB} and {\tt CosmoMC} are part of the {\tt FEARec++} distribution. Through this modular setup, extensions of {\tt FEARec++} to other variables are straightforward.

\vspace{-3mm}
\section{Recombination history eigenmodes}
\label{sec:recomb}
\begin{figure}
\centering
\includegraphics[width=\linewidth, trim={0 14 0 0},clip]{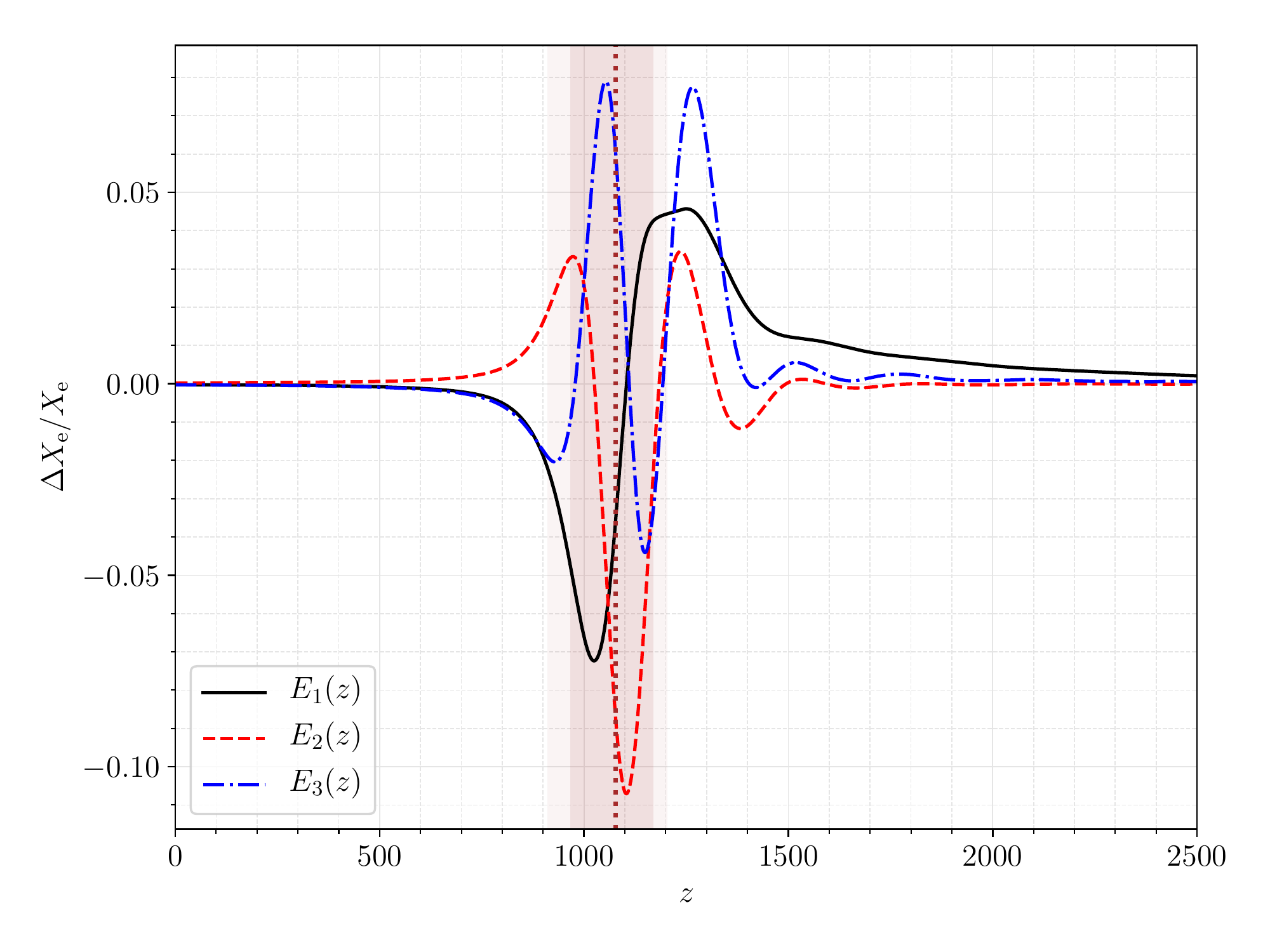}
\includegraphics[width=\linewidth, trim={0 20 0 0},clip]{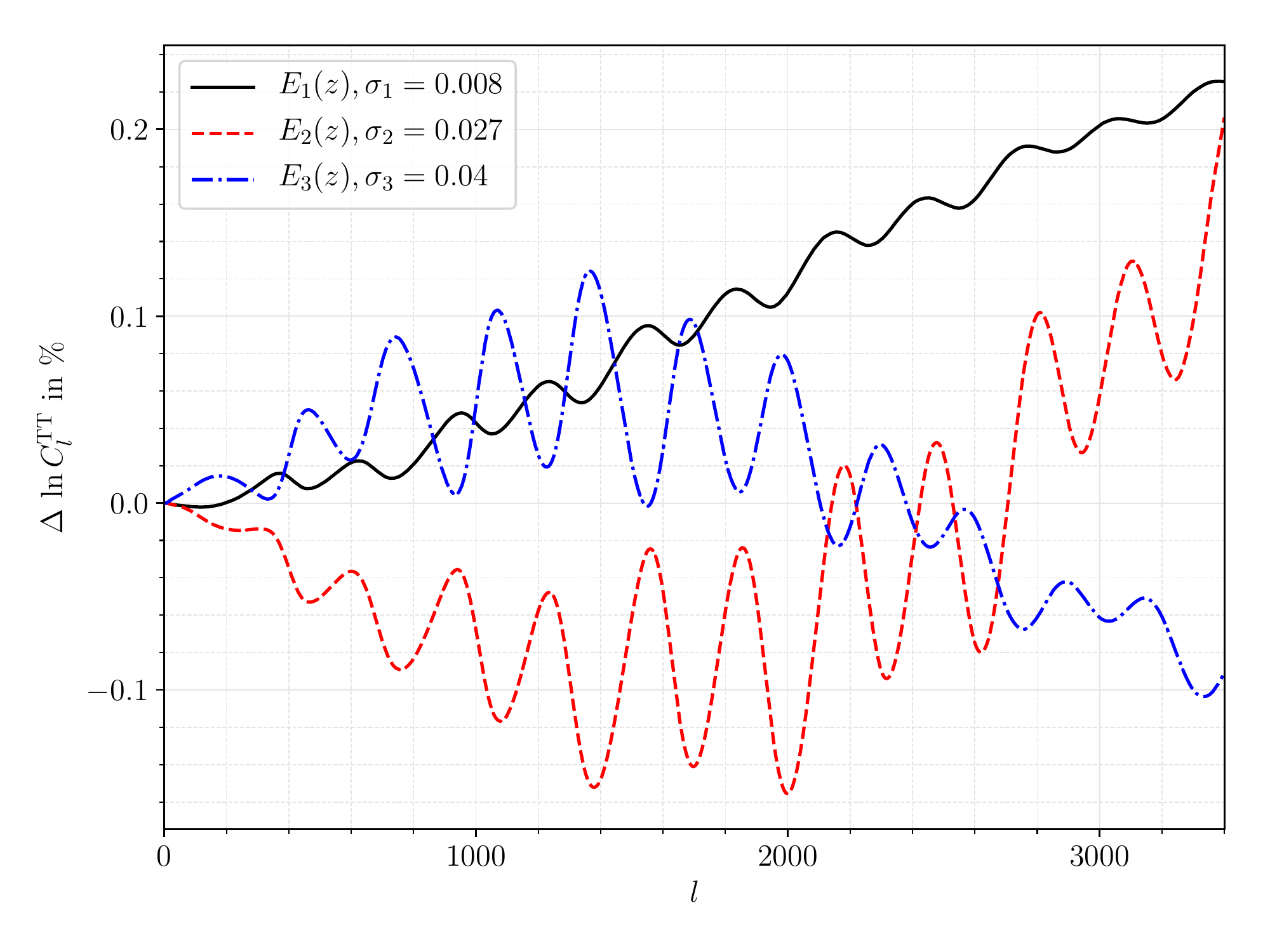}
\caption{{\it Top: } The three PCs emerging from the analytical Fisher method with a CVL experiment ($l_{\rm max} = 3500$). The functions are normalised with the coloured bands indicating the best fit value for the maximum of the Thomson visibility function, full width half and quarter maxima. The modes are consistent with \citet{Farhang2011}. {\it Bottom: } Differential CMB angular power spectrum due to the above principal components. The amplitudes of the eigenmodes that have generated these responses are weighted by their eigenvalues (EV) ($\sigma=1/\sqrt{{\rm EV}}$).}
\label{fig:cvl}
\end{figure}

\begin{figure}
  \centering
  \includegraphics[width=1.0\columnwidth, trim={0 14 0 0},clip]{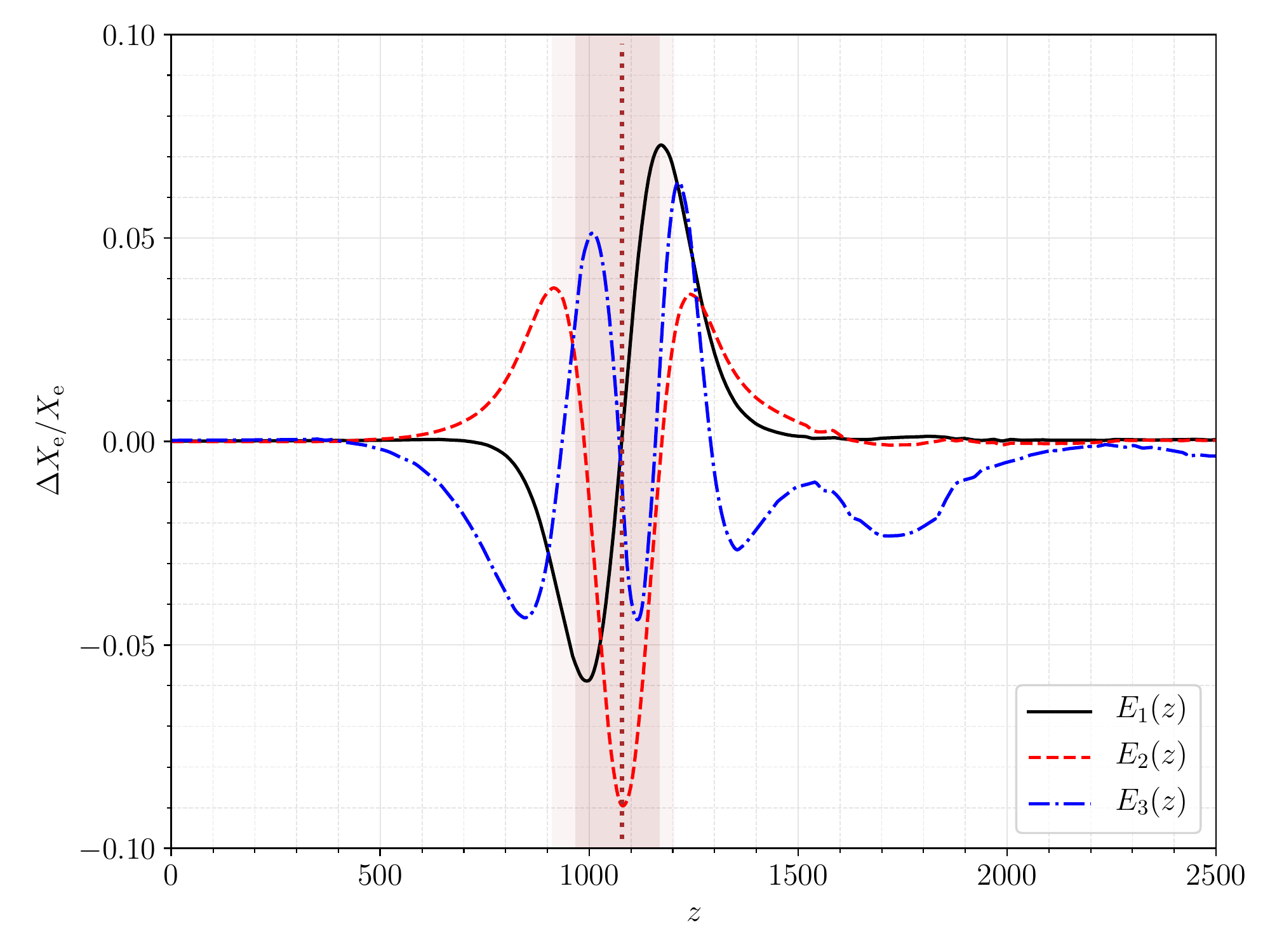}
  \includegraphics[width=1.0\columnwidth, trim={0 18 0 0},clip]{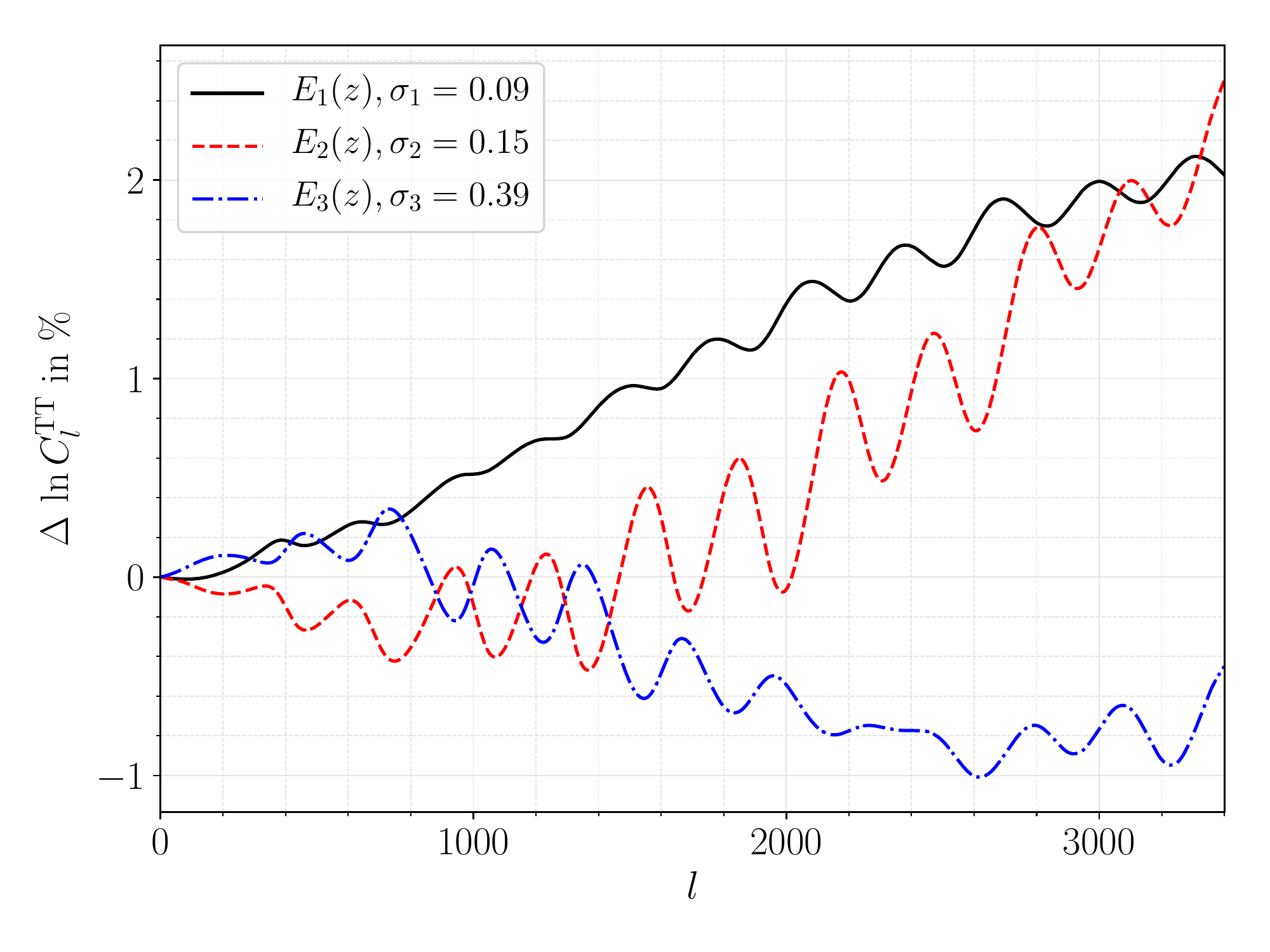}
  \caption{{\it Top:} PCs in the free electron fraction using the \planck data. The bands of the visibility function are identical to those in Fig.~\ref{fig:cvl}. {\it Bottom:} The relative differences emerging from these principal components in the CMB power spectrum. The mode scaling is the same as Fig.~\ref{fig:cvl}.}
  \label{fig:planck}
\end{figure}

In this section, we present the eigenmodes, referred to as principal components (PCs), in the free electron fraction $\xe$ from a CVL experiment shown in Fig.~\ref{fig:cvl} and those for the \planck 2015 data release in Fig.~\ref{fig:planck}. We apply the direct likelihood method explained in Sec.~\ref{sec:direct_method} and Appendix~\ref{app:direct} to the \planck 2015 data. This allows us to analyse the full likelihood dependence of standard parameters and generate highly uncorrelated eigenmodes.
We also briefly discuss the effect of mode-resolution, parameter marginalization and data combinations.

\subsection{Cosmic-variance limited modes}
\label{sec:recomb_cvl}
Firstly, in Fig.~\ref{fig:cvl} we present the PCs for a CVL experiment with no systematic noise component except a beam-limiting multipole truncations at $l_{\rm max} = 3500$. The PCs are ranked by their eigenvalues (i.e. ${\rm EV}_1>{\rm EV}_2$). The bands in the $\xe$ history ({\it top}) show the maxima, half maxima and quarter maxima of the Thomson visibility function. The responses in the CMB spectra ({\it bottom}) are scaled by their predicted errors such that $\sigma_i\sim1/\sqrt{{\rm EV}_i}$, all of which are shown in Table~\ref{tab:eigensolver} as well. The non-orthogonality of the first 3 CVL eigenmodes is $<10^{-7}$  for all combinations.

The first principal component $E_1$ generates a large tilt in the CMB spectrum shown in Fig.~\ref{fig:cvl}. The increase in the CMB temperature power spectra as a function of scale is reminiscent of tilting effects of the power spectrum synonymous with a change in the scalar power index, $\ns$. 
The second principal component, $E_2$ mainly leads to a shift in the position of the peaks. This is similar to the translational effects from varying the distance to the surface of last scattering, which mirrors the Hubble constant $H_0$ and the angular size of the last scattering surface $\theta_A$. 
\begin{table}
    \centering
    \begin{tabular}{l c c}
    \hline\hline
    Predicted errors & CVL experiment & \planck 2015 data  \\
    \hline
    $\sigma_1$ & $0.008$ & $0.088$\\
    $\sigma_2$ & $0.027$ & $0.147$\\
    $\sigma_3$ & $0.040$ & $0.394$\\
    \hline\hline
    \end{tabular}
    \caption{Esimated errors for the principal components from the eigensolver. These are the estimated errors given by the CVL modes (Fig.~\ref{fig:cvl}) and the \planck modes (Fig.~\ref{fig:planck}).} 
    \label{tab:eigensolver}
\end{table}
The third principal component is a combination of both effects with more oscillatory features. In Fig.~\ref{fig:cvl}, $E_3$ mirrors $E_2$ over a large number of multipoles, however there are subtle changes in spectral tilt at both extremes of the $\ell$ range. Higher order principal components display a larger number of these oscillations. Our findings agree well with previous analyses of a CVL experiment \citep{Farhang2011, Farhang2013}.

\subsection{\planck likelihood sampling modes}
\label{sec:recomb_plik}
Our reference \planck dataset is the \planck 2015 TTTEEE high-$\ell$ data alongside the lowTEB low-$\ell$ data \citep{Planck2015like}. Subsequently, we add \planck lensing and the SDSS DR12 BAO \citep{SDSSDR12} data. After applying the direct likelihood methods, including marginalisation over cosmological and standard parameters, we obtain the PCs shown in Fig.~\ref{fig:planck}. The estimated errors shown in Fig.~\ref{fig:planck} are $\simeq10$ times larger than the CVL equivalent as seen in Table~\ref{tab:eigensolver}. This arises from the added marginalisation and the resulting minor correlations between the eigenmodes. As we shall see in Sec.~\ref{sec:mcmc}, the estimated errors are indeed closely recovered in the full MCMC analysis, highlighting one of the improvements achieved with {\tt FEARec++}.
The first three \planck modes are furthermore orthogonal at the level of $<10^{-4}$.

Comparing with the CVL modes, the first principal component ($E_1$) loses the smoothed edge at the lower peak ($z\simeq 1200$) and also sharpens towards $z\simeq 1500$. The large-scale oscillatory features embedded in the CMB response from the first component are smeared out in the \planck case (cf. Fig.~\ref{fig:cvl} and \ref{fig:planck}).  
The second PC has a similar shape to the one found in Fig.~\ref{fig:cvl}, however some of the edge effects at $z\simeq 1400$ are removed for the modes in Fig.~\ref{fig:planck}. This leads to a slightly damped response in the CMB power spectrum when compared to Fig.~\ref{fig:cvl}. As with the CVL case, these changes are synonymous with the shifting of the angular size of the last scattering surface $\theta_A$.
In the third eigenmode, larger features arise at high redshifts ($1500<z<2000$). These extra features remove the mirroring behaviour between $E_2$ and $E_3$ from Fig.~\ref{fig:cvl}.

In our stability analysis it turned out to be instructive to directly study the Fisher matrix elements. A plot of the marginalised Fisher matrix is shown in Fig.~\ref{fig:planck_matrix}. The full details of the marginalisation discussion can be found in Appendix~\ref{app:marg}. The intermediate redshift range, $z\simeq 900-1300$, highlights the range of functions which are most constrained by the principal components in Fig.~\ref{fig:planck}. While the first component $E_1$ closely reflects the diagonal of the Fisher matrix, the second component $E_2$ appropriately reflects the opposite diagonal through the cross correlated elements. The Fisher elements decrease closer to $z\simeq 1100$, which is related to the dip in the first mode of Fig.~\ref{fig:cvl} and \ref{fig:planck} around the maxima of the Thomson visibility function.

\begin{figure}
  \centering
  \includegraphics[width=\linewidth]{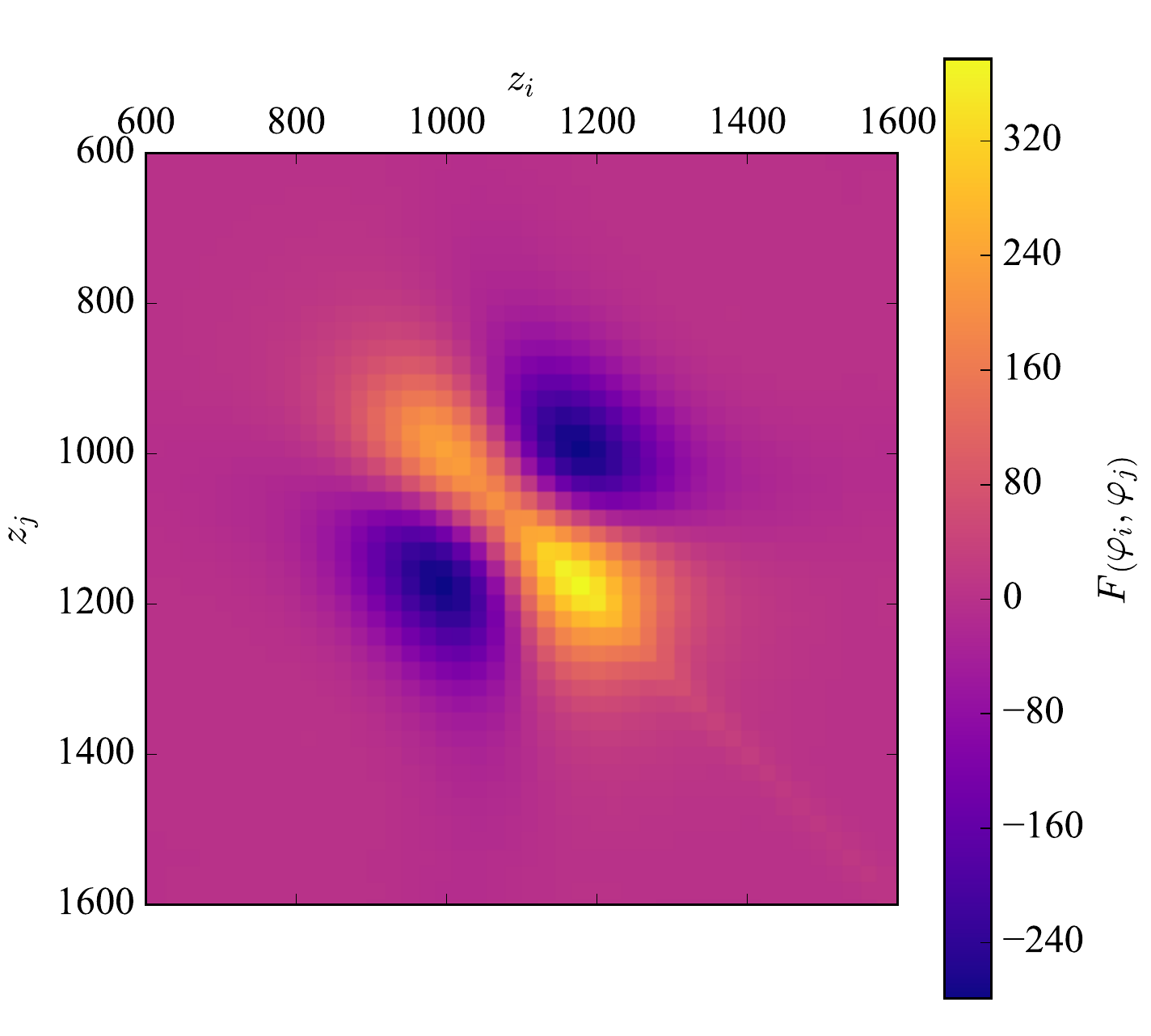}
  \caption{Marginalised matrix plot from the \planck data that led to the generation of the modes in Fig.~\ref{fig:planck}. The horizontal and vertical axes reference the redshifts of the basis functions, with the Fisher matrix elements being colour coded according to the key.}
  \label{fig:planck_matrix}
\end{figure}

\subsubsection{Studying the amplitudes of the basis functions}
\label{sec:amplitude}
\test{The eigenmodes are added to the fiducial ionization  history and translate from the recombination code to the Boltzmann code {\tt CAMB} efficiently, once we have dealt with the numerical issues (i.e. Appendix~\ref{app:dtau}).} However, the likelihood function has numerical limits. If the response from the basis function is too small, the response from the likelihood hits a given noise limit and perturbations in the tails, far away from the maxima of the Thomson visibility function, saturate by the numerical noise of the likelihood code. To counter this, we rescale the basis function $\varphi_i(z)$ is by the inverse Fisher elements from the CVL example, 
\begin{equation}
    \varphi_i'(z) = \frac{\varphi_i(z)}{\sqrt{F_{ii}}}.
    \label{eq:rescale_phi}
\end{equation}
When we use this setting, the $\ln\mathcal{L}$ responses for each basis function become comparable and we scale them by the amplitudes. This leads to clean, numerically stable eigenmodes. For various datasets, the CVL Fisher matrix diagonals are shown in Fig.~\ref{fig:diagonals}.

\vspace{-3mm}
\subsection{Convergence of directly sampled eigenmodes}
\label{sec:converge}
\begin{figure}
  \centering
  \includegraphics[width=\linewidth, trim={0 35 0 0},clip]{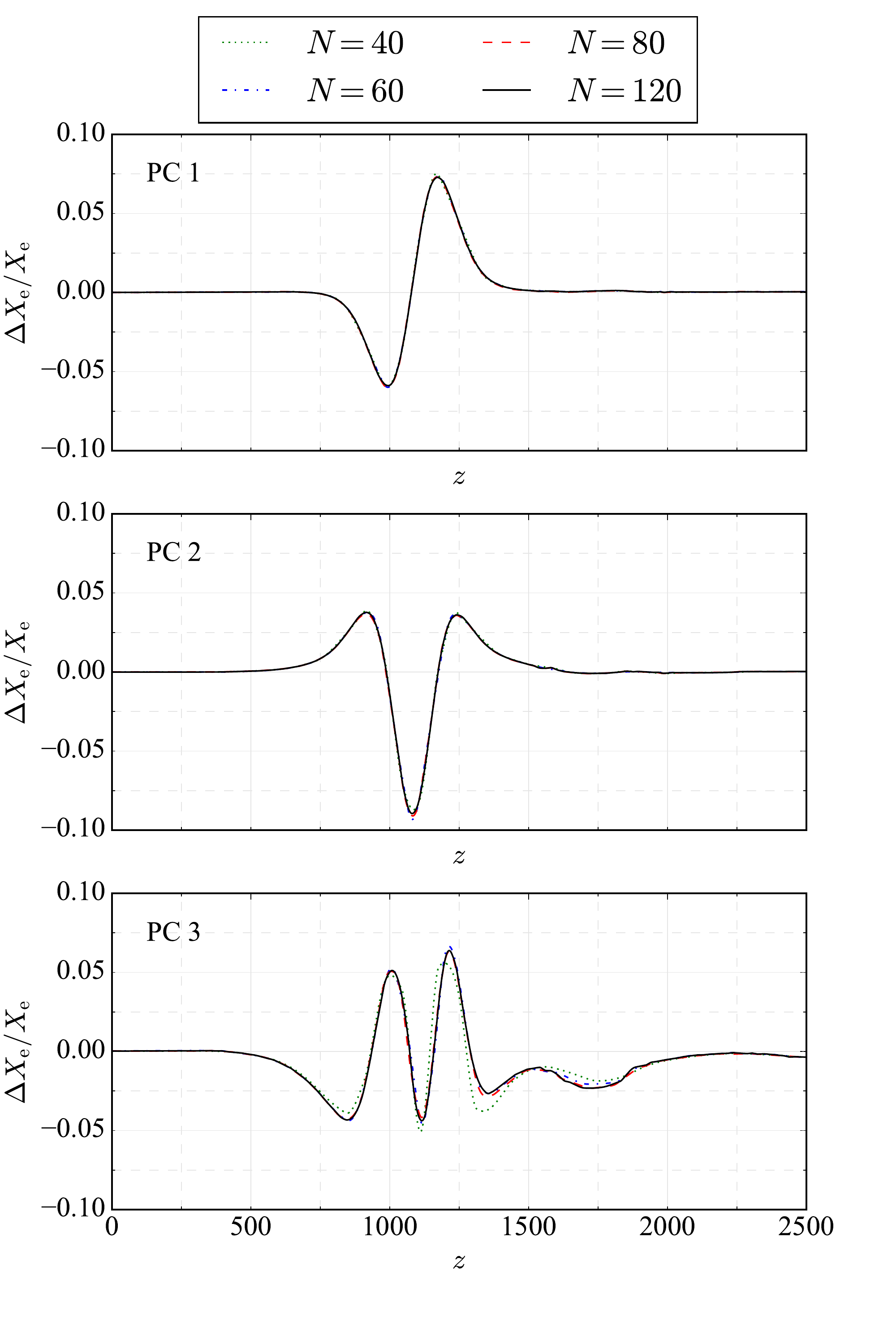}
  \caption{\planck eigenmodes for varying numbers of basis functions. The first three marginalised principal components are shown for the different basis levels of $N=\{40,60,80,120\}$. The resolution of the basis functions is $\Delta z \simeq 22$ for $N=120$ with the basis width defined in Appendix~\ref{app:basis}.}
  \label{fig:number_test_plik}
\end{figure}

In this section, we test the convergence of the directly sampled components using \planck data. We show the changes in the eigenmodes as a function of the grid resolution we obtained in Fig.~\ref{fig:number_test_plik}. The increased resolution isolates the finer features and also removes the spurious mode correlations created by the interpolation routines. 
The eigenmodes shown in Fig.~\ref{fig:number_test_plik} become indistinguishable after $N>60$. However, inter-mode correlations still reach the $\simeq1\%$ limit before marginalisation. It is only when we use $N=120$ that these non-orthogonalities disappear. 

To remove all non-orthogonalities to a higher precision (we shall discuss why this is important in Sec.~\ref{sec:mcmc}), we attempted a finer resolution grid. The rescaling of the basis amplitudes discussed in Sec.~\ref{sec:amplitude} is affected by the widths of the functions as well. However, when we increased the resolution for $N=160$, the basis function responses hit the numerical limit of the likelihood code. Further increasing the amplitude of the functions near the Thomson visibility function, caused highly non-linear responses and the obtained modes were no longer smooth. 

A variable grid with higher resolution around $z\simeq 1100$ was also studied. In this case, we set the effective number of basis functions to,
\begin{equation}
 N_{\rm res}=\begin{cases}
    160, & \text{$900<z<1300$}.\\
    120, & \text{otherwise}.
 \end{cases}
\end{equation}
However, this left us with large non-orthogonalities ($\simeq10\%$) after interpolation. In order to smooth the noisiness at high redshifts, we furthermore applied a Gaussian filter to the eigenvectors; however this also added non-orthogonalities, once done too aggressively. 

After all our optimizations, the eigenmodes presented in Fig.~\ref{fig:planck} were the best we could obtain with the current machinery. The main limitation was found to stem from the numerical precision of the likelihood code (see Appendix~\ref{app:direct}). However, the results shown in Sec.~\ref{sec:mcmc} significantly build our confidence surrounding these modes. 
Further improvements may be possible with the new \planck 2018 likelihood code, but we leave this exploration to future work.

\subsection{Effect of parameter marginalisation} 
\label{sec:param_marg}
\begin{figure}
  \centering
  \includegraphics[width=\linewidth, trim={0 40 0 0},clip]{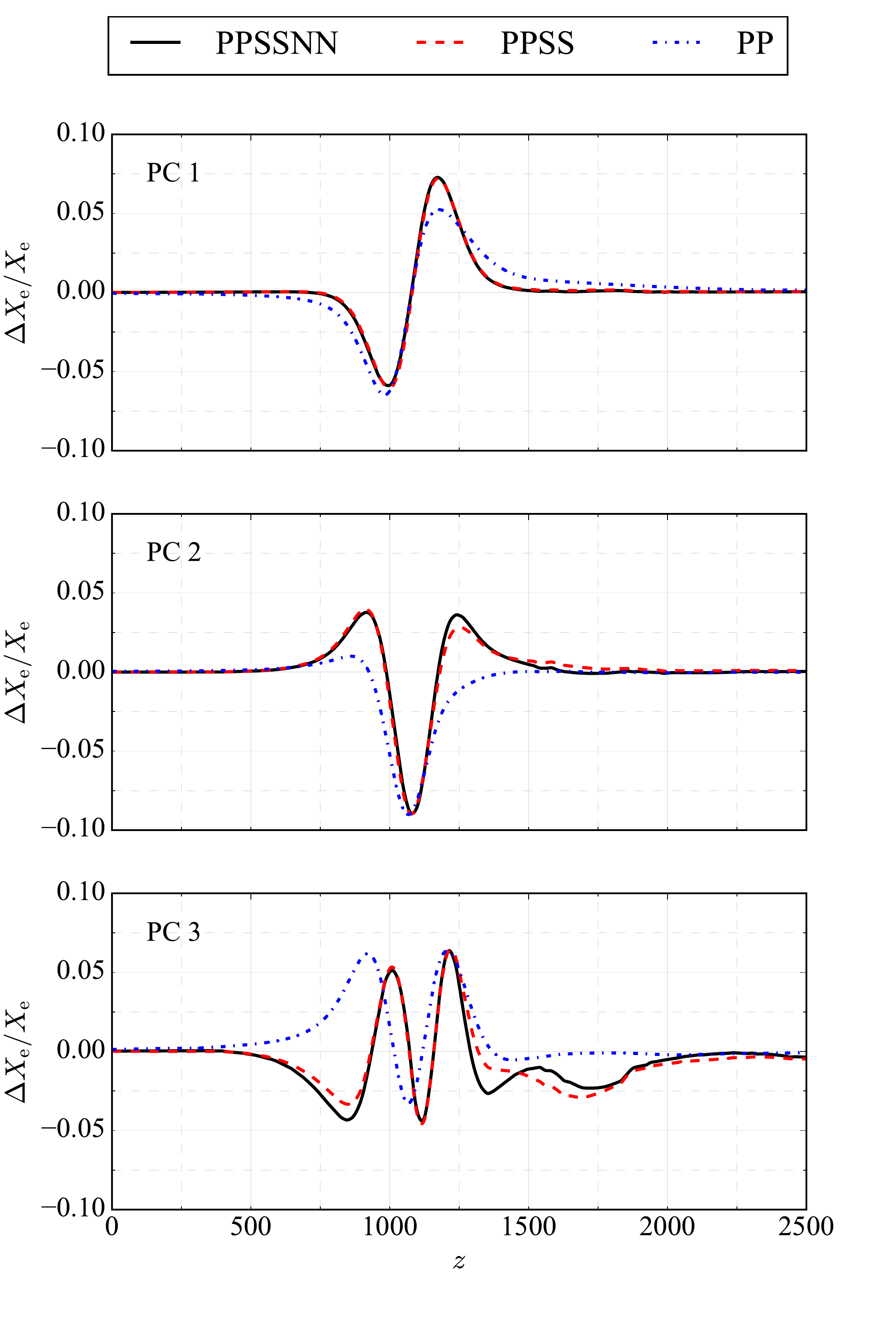}
  \caption{Changes in the first three principal components for different levels of marginalisations. The black line (PPSSNN) includes marginalisation over the full cosmological parameters and nuisance parameters as shown in Fig.~\ref{fig:planck}. The red dashed line (PPSS) does not include the nuisance parameters. The blue bashed lines (PP) includes just the perturbation components, with no marginalisation.}
  \label{fig:ppssnn}
\end{figure}

In this section, we investigate how the eigenmodes change as we remove the nuisance and standard parameter marginalisation. 
This is achieved by omitting the corresponding blocks in the Fisher information matrix before the PC construction. The results are compared in Fig.~\ref{fig:ppssnn}.

There are negligible changes in the first eigenmode when we neglect nuisance parameter marginalisation. The perturbation-only modes (PP) have a longer tail at high redshifts.
We also find a relatively minor effect due to marginalisation over nuisance parameters for the second eigenmode. However, removing standard parameter marginalisation has a pronounced effect, removing the positive turns at $z\simeq900$ and $z\simeq1300$. 
For the third mode, both the standard and nuisance parameter marginalisation have a significant effect on its shape. Without any marginalisation, the mode is more focussed around the hydrogen recombination era, with only weak tails into the neutral helium recombination era ($z\simeq 1500-2000$). In this case, the third mode becomes quite similar to the corresponding CVL eigenmode in Fig.~\ref{fig:cvl}. When marginalising over nuisance parameters, there is enhanced small-scale noise at high redshifts. This indicates further optimization is possible; however we did not investigate this as our eigenmodes are already convincingly decorrelated from the \planck parameters. 

\vspace{-3mm}
\subsection{Adding lensing and BAO data}
\label{sec:data_marg} 
\begin{figure}
  \centering
  \includegraphics[width=\linewidth, trim={0 35 0 0},clip]{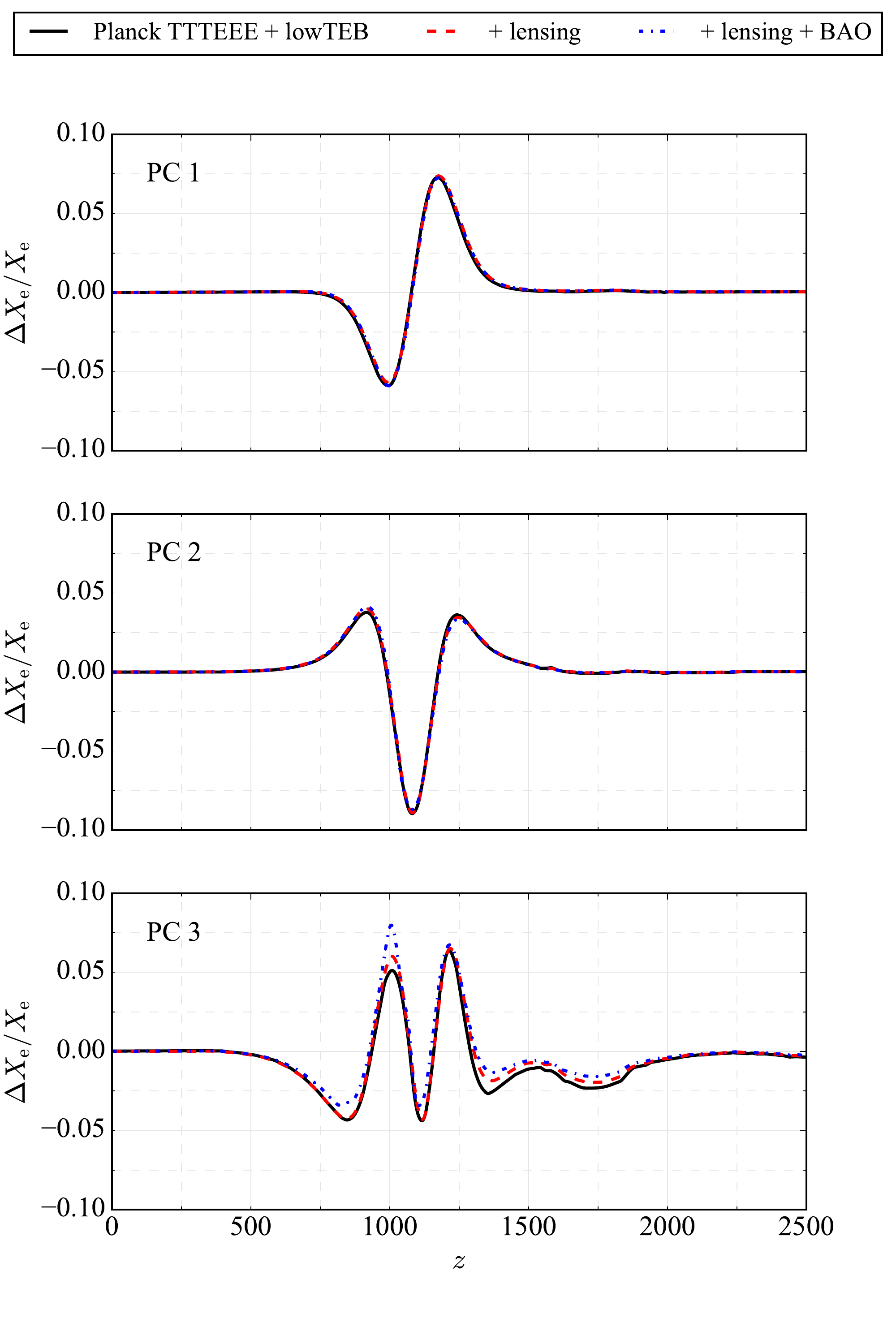}
  \caption{Comparisons of the first 3 principal components for different datasets. The reference case (black, solid line) includes \planck 2015 data for high l, (Planck TTTEEE) and low-$\ell$ (lowTEB) temperature and polarisation. Then, we include \planck lensing from 2015 and also another case where BAO data is included. All modes are normalised as $\int E^2_i(z)\,\id z=1$.}
  \label{fig:datasets}
\end{figure}

\begin{figure}
  \centering
  \includegraphics[width=1.01\linewidth, trim={4 10 0 0},clip]{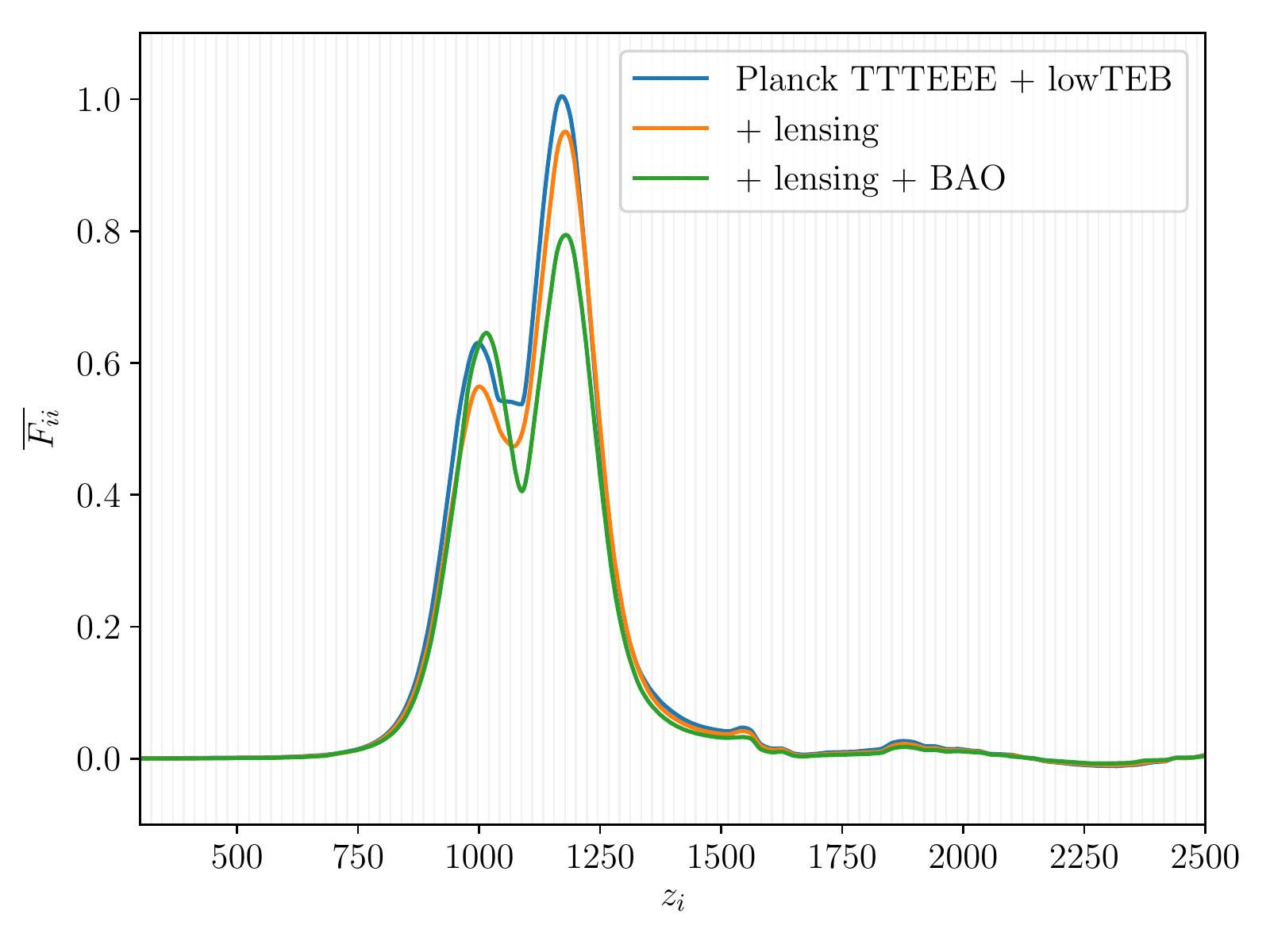}
  \caption{The diagonal elements from the different Fisher matrices for the PCs presented in Fig.~\ref{fig:datasets}. The different Fisher diagonals have been normalised by their total area. The grid lines (grey) show the resolution of the functions chosen in the basis set with $N=120$.}
  \label{fig:diagonals}
\end{figure}

The effects of adding extra datasets when we create the eigenmodes are shown in Fig.~\ref{fig:datasets}. We compare the standard \planck data combination against cases where we add CMB lensing data \citep{PlanckLensing2016} and BAO data \citep{SDSSDR12} as well.
The majority of the eigenmode structures are kept fixed as more datasets are added; however in the third eigenmode of Fig.~\ref{fig:datasets}, exhibits a larger spike at $z\simeq 1000$. This dilutes the higher redshift amplitude parts at $z\gsim1300$. When we look at the diagonals of the respective Fisher matrices in Fig.~\ref{fig:diagonals}, we can see this redistribution between the peaks at $z\simeq1000$ and $z\simeq1200$ more clearly. Note that the grey lines on this figure replicate the finite number of eigenvector points that are evaluated and interpolated between (as described in Appendix~\ref{app:interpolation}).

\vspace{-4mm}
\subsection{Comparing orthogonalities to previous analysis}
\label{sec:planck_comp}
In this section, we compare our principal components to \planck 2015 mode analysis \citep{Planck2015params}. The correlations $\xi_{ij}$ between each eigenmode are shown in Table~\ref{tab:correlations}.
Using the same data combination, we present the orthogonalities between the eigenmodes in Table~\ref{tab:correlations}, where $\xi_{ij}$ is the correlation between modes $i$ and $j$. In Table~\ref{tab:correlations}, the  non-orthogonality between our eigenmodes is at least $\simeq 1000$ times smaller than that of the previous work \citep{Planck2015params}. 
As discussed in Sec.~\ref{sec:converge}, smoothing of the modes is likely one of the causes of these relatively large mode projections for the original \planck 2015 modes. Our interpolation scheme avoids this issue and thus improves the performance of our modes in the analysis and the direct parameter projection method explained in Sect.~\ref{sec:proj}.

\begin{table}
\begin{center}
\begin{tabular}{ c | c | c }
\hline\hline
Correlation & \planck 2015 & HC 2019 \\
\hline
$\xi_{21}$ & -0.05 & $2\times10^{-6}$ \\
\hline
$\xi_{31}$ & -0.09 & $1\times10^{-4}$ \\
\hline
$\xi_{32}$ & -0.07 & $4\times10^{-5}$ \\
\hline\hline
\end{tabular}
\end{center}
\caption{The correlations between the first three eigenmodes from the previous \planck 2015 analysis \citep{Planck2015params} and the ones presented in Fig.~\ref{fig:planck}. The eigenmodes have been constrained for the same given dataset (\planck + lensing + BAO). Our modes generated with \planck and \planck + lensing show even better orthogonality.}
\label{tab:correlations}
\end{table}

\section{Constraining mode amplitudes with data}
\label{sec:mcmc}
The obtained PCs for the \planck data, presented in Fig.~\ref{fig:planck}, are added as a perturbation to the recombination history with an amplitude $\mu_i$. These amplitudes are allowed to vary in an MCMC simulation and we vary for the normal set of \LCDM parameters ($\{\omb,\omc,\theta,\tau,\ns,\logA\}$ and nuisance parameters. The results for the first three modes are shown in Table~\ref{tab:planck}.

\begin{figure}
    \centering
    \includegraphics[width=0.98\linewidth, trim={0 5 0 0},clip]{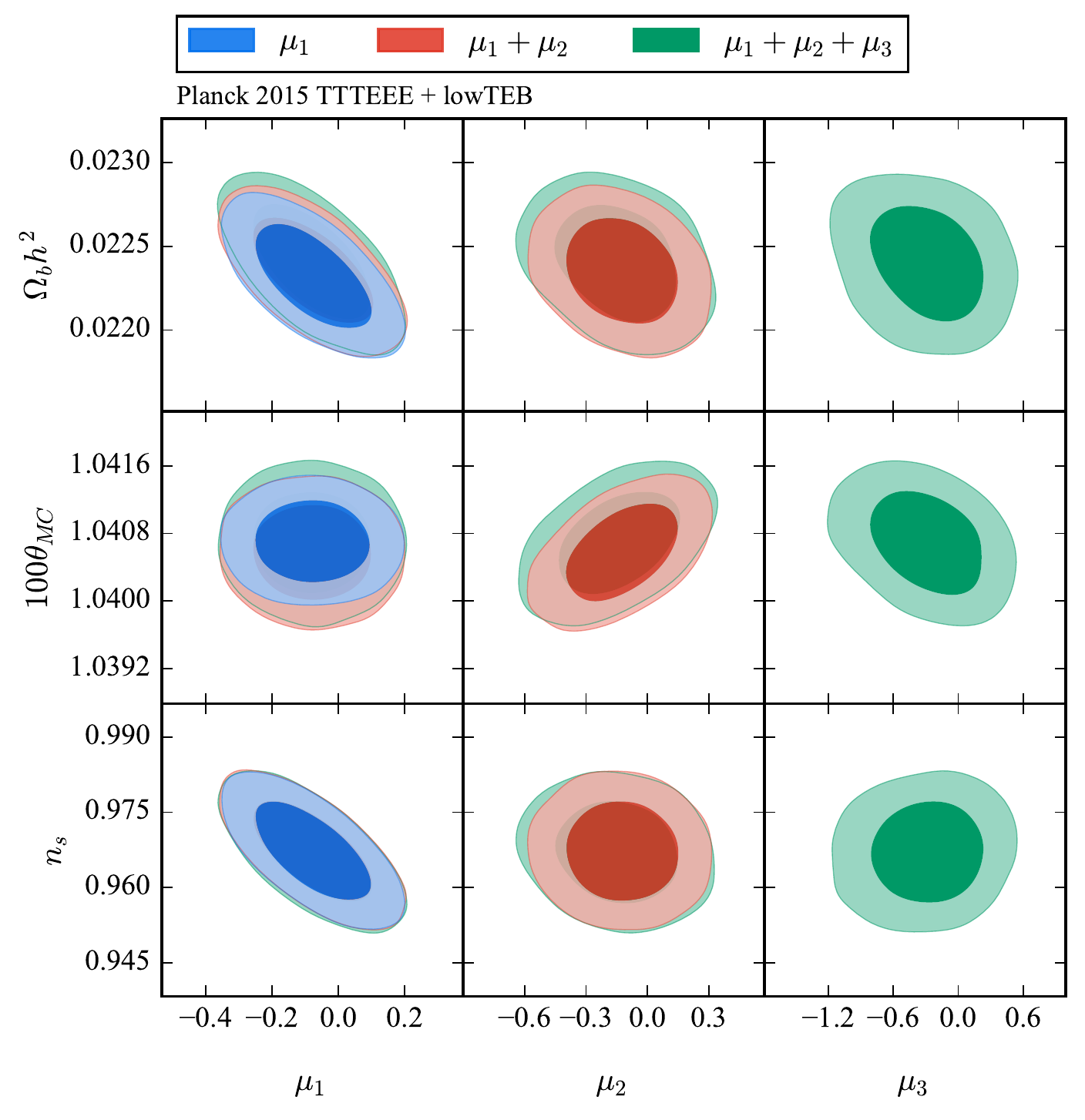}
    \caption{Posterior contours illustrating the key degeneracies between the standard $\Lambda$CDM parameters and the recombination eigenmode amplitudes. We have omitted $\left\{\omc,\tau,\logA\right\}$ because they are largely unaffected by the recombination eigenmodes. [See Fig.~\ref{fig:contours_extra_planck} for those contours.]}
    \label{fig:contours_degen}
\end{figure}

For the PCs included in Table~\ref{tab:planck}, we find $|\mu_i|/\sigma_i<1$, meaning that standard recombination is favoured. However, some of the standard parameters do move around slightly when the eigenmodes are added into the analysis. The parameter degeneracies are displayed in the posterior contours shown in Fig.~\ref{fig:contours_degen}.
Both $\omb$ and $\omc$ only drift away from their initial value by $\simeq0.5\sigma$. The angular size of the last scattering surface $\thetaMC$ varies a small amount when we add the eigenmodes; however, we notice that when the third eigenmode is added the value of $\theta$ is closer to the fiducial value before the eigenmodes were added, with a slightly higher error ($\sigma_\theta \simeq 1.2\sigma_\theta^{\rm fid}$). The majority of this error increase is driven by $E_2$. The primordial power spectrum amplitude $\As$ and the reionization optical depth remain unaffected, however the degeneracy between $\ns$ and $\omb$ drives the error of $\ns$ up as well. This is due to the tilting effect on the CMB power spectra from $E_1$ which is reminiscent of $\ns$ effects as pointed out in Sec.~\ref{sec:recomb_cvl}. 

\begin{figure}
    \centering
    \includegraphics[width=\linewidth]{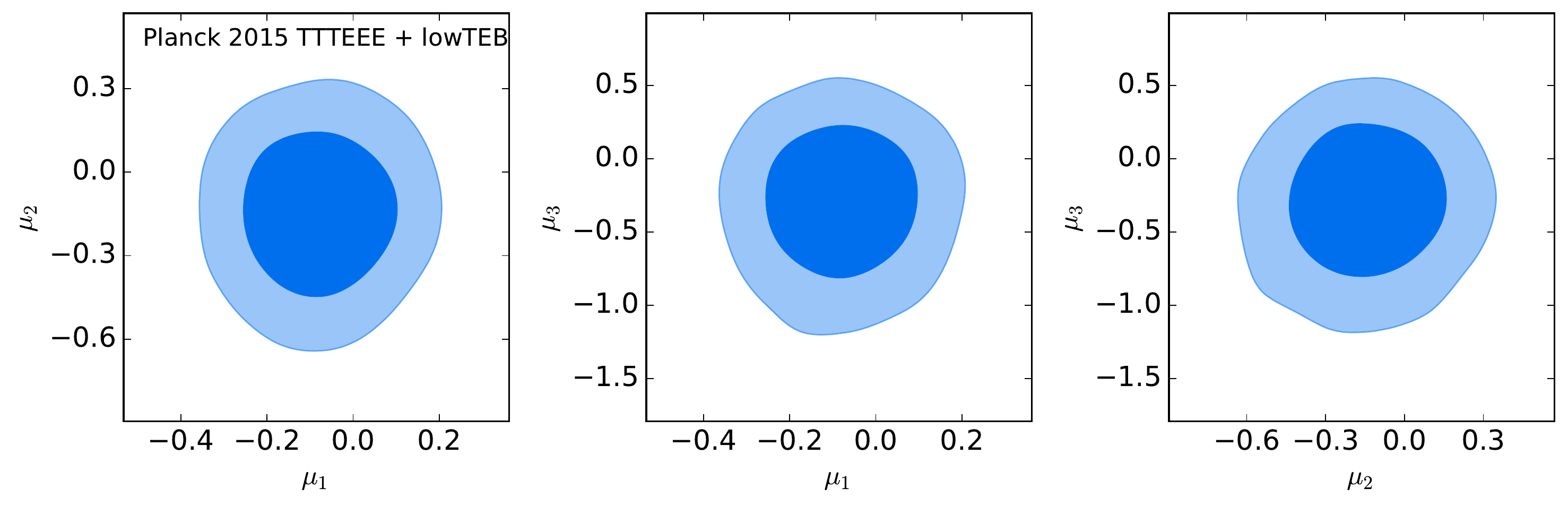}
    \caption{Posterior contours between the first three PCs when they are added to the standard parameters in an MCMC simulation. These have been generated using the \planck TTTEEE + lowTEB dataset.}
    \label{fig:contours_mu}
\end{figure}

\begin{figure}
    \centering
    \includegraphics[width=0.98\linewidth, trim={0 5 0 0},clip]{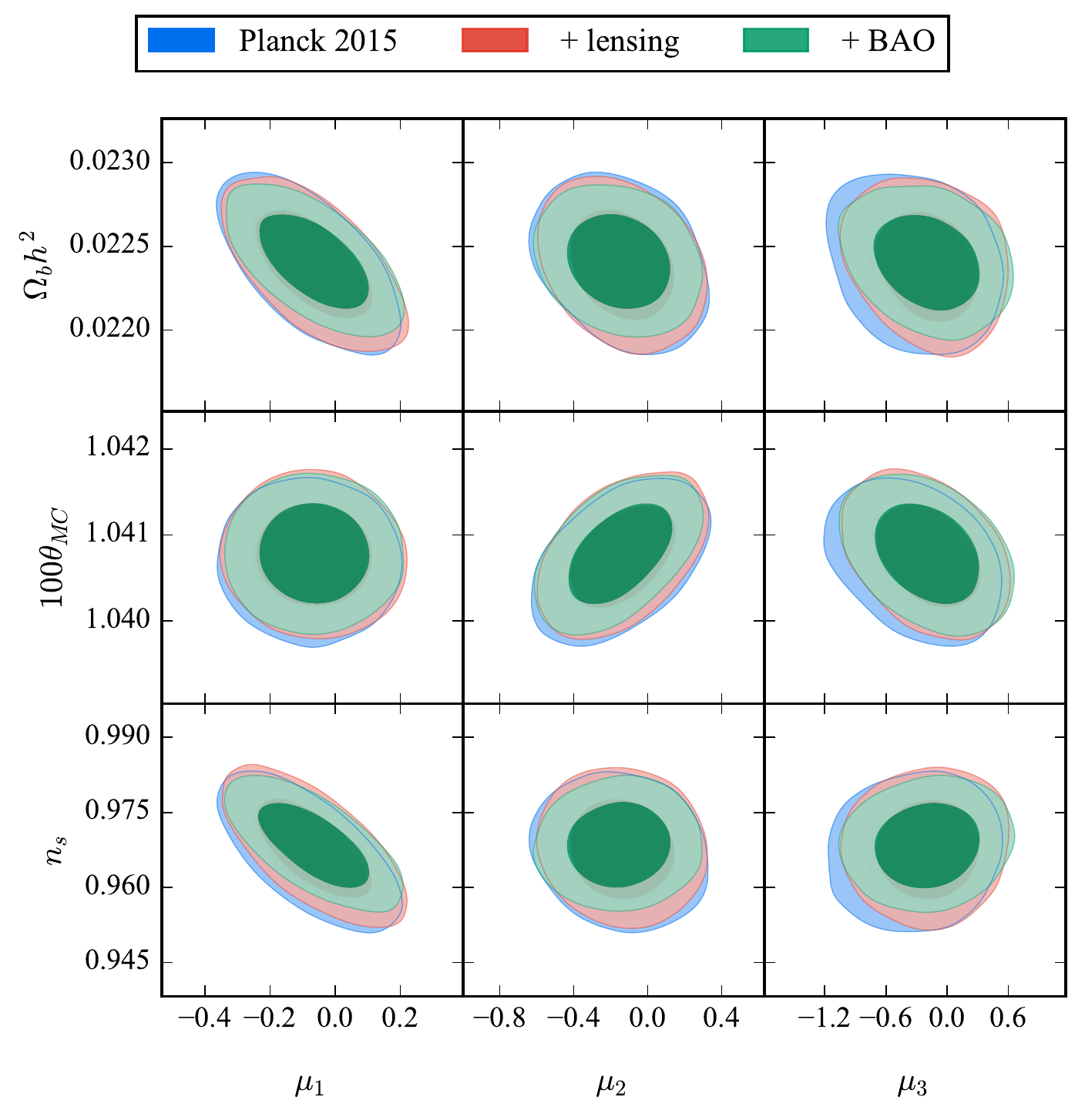}
    \caption{Posterior contours for the degenerate parameters comparing the different datasets with their respective constrained modes. Here we have added the lensing likelihood as well as the {\it BOSS} and {\it SDSS} BAO data. Here we have used the Markov chains for marginalising over all three eigenmode amplitudes, with the modes being sourced from the \planck TTTEEE + lowTEB combination.}
    \label{fig:contours_data}
\end{figure}

As we add the eigenmodes, the variations in the standard parameters are  comparable to those found in the previous work \citep{Planck2015params}.
The main difference in this analysis is the orthogonality of the eigenmodes. This becomes quite apparent when we compare the errors on the eigenmodes. Whilst the first two principal components are comparable to the previous analysis, the error on the third eigenmode is $\simeq2.5$ times better in this work. We believe this is mainly due to the interpolation scheme we used to generate the eigenmodes from the resultant eigenvectors (see Appendix~\ref{app:interpolation} for more details). For full clarity, we present the posterior contours between the eigenmodes in Fig.~\ref{fig:contours_mu} to highlight the lack of $\mu_i$ correlations. 
Furthermore, as shown in Table~\ref{tab:eigensolver}, the results from this MCMC simulation are quite consistent with the errors predicted from the eigensolver when the mode was generated. For example, the third eigenmode has an predicted error of $\sigma_3\simeq 0.39$, whilst the full MCMC yielded $\sigma_3\simeq0.35$.

\begin{table*}
    \centering
    \begin{tabular} { l  c c c c}
\hline\hline
Parameter & Planck TTTEEE \& lowTEB & + 1 mode & + 2 modes & + 3 modes\\
\hline
$\Omega_b h^2$ & $0.02224\pm 0.00016$ & $0.02232\pm 0.00020$ & $0.02235\pm 0.00021$ & $0.02240\pm 0.00022$\\[0.5mm]
$\Omega_c h^2$ & $0.1198\pm 0.0015$ & $0.1197\pm 0.0015$ & $0.1194\pm 0.0016$ & $0.1193\pm0.0016$\\[0.5mm]
$100\theta_{MC}$ & $1.04073\pm 0.00033$ & $1.04071\pm 0.00032$ & $1.04058\pm 0.00038$ & $1.04070\pm 0.00040$\\[0.5mm]
$\tau$ & $0.080\pm 0.017$ & $0.082\pm 0.018$ & $0.084\pm 0.018$ &  $0.087\pm 0.018$\\[0.5mm]
${\rm{ln}}(10^{10} \As)$ & $3.095\pm 0.032$ & $3.099\pm 0.035$ & $3.102\pm 0.035$ & $3.109\pm 0.035$\\[0.5mm]
$n_s$ & $0.9652\pm 0.0048$ & $0.9671\pm 0.0064$ & $0.9672\pm 0.0065$ & $0.9672\pm 0.0067$\\
\hline
$\mu_1$ & $--$ & $-0.08\pm 0.11$ & $-0.08\pm 0.12$ & $-0.08\pm 0.12$\\[0.5mm]
$\mu_2$ & $--$ & $--$ & $-0.13\pm 0.18$ & $-0.14\pm 0.19$\\[0.5mm]
$\mu_3$ & $--$ & $--$ & $--$ & $-0.30\pm0.35$\\
\hline
$H_0$ & $67.25\pm 0.65$ & $67.35\pm 0.68$ & $67.45\pm 0.71$ & $67.54\pm 0.74$\\
\hline\hline
\end{tabular}
    \caption{Constraints from the converged eigenmodes using \planck 2015 data. The standard \planck constraints with the standard recombination picture is included alongside the limits generated from adding 1,2 and 3 principal components. The limits also show the changes in the standard $\Lambda$CDM cosmological parameters as described in Sec.~\ref{sec:formalism}. All errors quoted are $68\% $ limit errors and here we have also included the marginalised value derived for the Hubble constant.}
    \label{tab:planck}
\end{table*}

\begin{table*}
    \centering
    \begin{tabular} { l  c c c}
\hline\hline
Parameters & \planck 2015 TTTEEE + lowTEB &  + lensing &  + lensing + BAO\\
 & (3 modes) & & \\
\hline
$\Omega_b h^2$ & $0.02240\pm 0.00022$ & $0.02239\pm 0.00022$ & $0.02241\pm 0.00019$\\[0.5mm]
$\Omega_c h^2$ & $0.1193\pm0.0016$ & $0.1187\pm 0.0016$ & $0.1183\pm 0.0011$\\[0.5mm]
$100\theta_{MC} $ & $1.04070\pm 0.00040$ & $1.04077\pm 0.00040$ & $1.04079\pm 0.00038$\\[0.5mm]
$\tau$ & $0.087\pm 0.018$ & $0.068\pm 0.015$ & $0.070\pm 0.013$\\[0.5mm]
${\rm{ln}}(10^{10} \As)$ & $3.109\pm 0.035$ & $3.067\pm 0.027$ & $3.071\pm 0.024$\\[0.5mm]
$n_s$ & $0.9672\pm 0.0067$ & $0.9678\pm 0.0066$ & $0.9686\pm 0.0056$\\
\hline
{$\mu_1$} & $-0.08\pm 0.12$ & $-0.07\pm 0.12$ & $-0.06\pm 0.11$\\[0.5mm]
{$\mu_2$} & $-0.14\pm 0.19$ & $-0.14\pm 0.19$ & $-0.16\pm 0.19$\\[0.5mm]
{$\mu_3$} & $-0.30\pm0.35$ & $-0.19\pm 0.34$ & $-0.19\pm 0.35$\\
\hline
$H_0$ & $67.54\pm 0.74$ & $67.78\pm 0.71$ & $67.93\pm 0.50$\\[0.5mm]
\hline\hline
\end{tabular}
    \caption{Constraints from the three eigenmodes generated with \planck TTTEEE and lowTEB data. These are marginalised limits sampled from standard \planck data along with lensing and BAO data from {\it SDSS} (DR12) and {\it BOSS}. The parameters and error limits are the same as those in Table~\ref{tab:planck}.}
    \label{tab:planck_data}
\end{table*}

\subsection{Marginalised results for external datasets}
\label{sec:mcmc_data}
In this section, we focus on how the standard parameters and eigenmode amplitudes are affected as we add external data. The added datasets are consistent with those in Sec.~\ref{sec:data_marg}. 
We present the marginalised constraints for this comparison in Table~\ref{tab:planck_data}. It is important to note that these modes were generated with our reference \planck dataset (high-$\ell$ TTTEEE and low-$\ell$ TEB) and we have just added the extra datasets into the MCMC simulation. A similar procedure was used previously \citep{Planck2015params, Planck2018params}.
The final constraints and mode-correlation are indeed very similar across the data combinations, except that we have a reduction of the $\As$ error from adding lensing data and a lower errors for $\omb$, $\omc$ and $\ns$ when we add BAO data. These contractions are also visible in the posteriors of Fig.~\ref{fig:contours_data}. 

When we replicate the analysis for eigenmodes optimised for \planck, CMB lensing and BAO data, the results for $E_1$ and $E_2$ remain the same. However adding the third eigenmode ($E_3$) gives us worse errors than the constraints for the \planck-only modes. This arises from complications in the cross correlations of the Fisher matrix ($F_{ps}$). One could improve the numerical stability for these additional datasets; however, it is clear from Table~\ref{tab:planck_data} that the MCMC results are largely unaffected when we add CMB lensing and BAO, such that we did not explore this any further.

The results from the MCMC analysis allow us to test the standard recombination picture with the most-likely variations given the principal component analysis. Although a non-standard recombination scenario is disfavoured, given the marginalised errors, we can conclude that at the current level of precision the eigenmode results depend only marginally on the dataset. We can also conclude that the formalism, used with the machinery in {\tt FEARec++} generates highly orthogonal eigenmodes. \test{Given that the eigenmodes now include standard parameter marginalisation}, we can apply the results from the MCMC study in a new `{\it direct projection}' formulation, as outlined in the following section.

\section{Direct projections using the eigenmodes}
\label{sec:proj}
The model-independent results for the mode amplitudes can be directly used to obtain estimates for specific physical scenarios that modify the recombination history. Here, we introduce the method of directly projecting model-independent principal components of one physical variable (i.e. $\xe$) onto a parametrised variation in the same variable for a physical scenario. This allows us to derive parameter constraints associated with these parametrised variations without the need to rerun the MCMC analysis.

\subsection{Projection definition}\label{sec:proj_def}
Let us rewrite $\xe$ in terms of the principal components $E_i$ that we have found in this analysis. Formally we can express this as, 
\begin{equation}
    \ddxe(z) = \sum_i\rho_i\,E_i(z), \qquad
    \rho_i = \int\ddxe(z)\cdot E_i(z) \id z,
\label{eq:rho}
\end{equation}
where $\rho_i$ denotes the projection of the model-independent eigenmodes $E_i$ onto the parametrised physical variation $\Delta\xe$. Note we can only define the projections like this if the eigenmodes are orthogonal to a sufficient precision (e.g., $<1\%$) recreating an orthonormal set. This was not the case for the modes previously used for the \planck 2015 analysis (e.g., see Table~\ref{tab:correlations}).

The next step is to compare the $\rho_i$ to the obtained eigenmodes amplitudes from the MCMC. These refer to the $\mu_i$ from Table~\ref{tab:planck} and Fig.~\ref{fig:contours_mu}. Using the projections from the eigenmodes, their covariance matrix and the marginalised values from the Markov chain analysis, we can formulate an effective $\chi$-squared, 
\begin{equation}\label{eq:chi2}
     \chi^2 = \left(\Vec{\rho_i}-\Vec{\mu_i}\right)^{\rm T}\cdot\hat{\Sigma}^{-1}_{ij}\cdot\left(\Vec{\rho_i}-\Vec{\mu_i}\right),
 \end{equation}
where $\Vec{\rho}$ and $\Vec{\mu}$ are the projection and constrained amplitude vectors respectively. The matrix $\hat{\Sigma}$ is the covariance matrix between each of the marginalised eigenmode amplitudes, which can be obtained from the MCMC runs. \test{The direct projections use the same number of added eignemode amplitudes for their MCMC runs respectively. For example, the results for $\mu_1+\mu_2$ in Fig.~\ref{fig:vertical_projections}, are derived from marginalising over just the first two amplitudes in the MCMC.}
It is important to stress that this is just an effective $\chi^2$-analysis, which defines how well the projection of the parametrised variation $\rho_i$ fits onto the marginalised PC amplitudes $\mu_i$ given their uncertainties. Ideally, the mode covariance matrix, $\hat{\Sigma}(\mu_i\,\mu_j) = {\rm Diag}\left(\sigma^2_{i}\right)$, where $\sigma_i$ is the standard deviation of the eigenmodes for a given dataset; however, minor correlations can affect parameter constraints at the level of $\simeq1\%$. 
 
 The next step is to compare the effective strength of the model-independent variation with the physical/parametrised variations that we wish to measure. For this we must consider a simple linear scaling in the physical variation. We focus on a generic variation in some function $\xe$ due to a change in parameter $\parA$. For small changes in $\parA$, it is sensible to assume that,
\begin{equation}\label{eq:linear_proj}
     \ddxe(\Delta\parA,z) \propto \ddA.
 \end{equation}
 Therefore, we can carry this relation through our definition of the projection in Eq.~\eqref{eq:rho} such that,
 \begin{equation}\label{eq:rho_new}
     \rho_i\rightarrow\rho_{i,0}\left(\frac{\delA}{\delA_0}\right)= \hat{\rho}_i\;\Delta \parA,
 \end{equation}
 where $\delA_0$ is a reference change in a physical parameter to generate the parametrised change discussed previously. Here our new definition, $\hat{\rho}_i$ is a projection that has been normalised by the fiducial change that caused it. Using the linear relation from Eq.~\eqref{eq:linear_proj}, we now have a weighted projection variable that roughly scales the true projection by the $\delA$ that alters the functional variable $\Phi$ we are tracing. 
 This is basically a first order Taylor series expansion of the problem and allows us to estimate the $\chi^2$-changes. 
  
\vspace{-3mm}
\subsubsection{Mode-projections for various physical scenarios}
\label{sec:rho_i_results}
\begin{figure}
  \centering
  \includegraphics[width=1.0\linewidth, trim={0 10 0 0},clip]{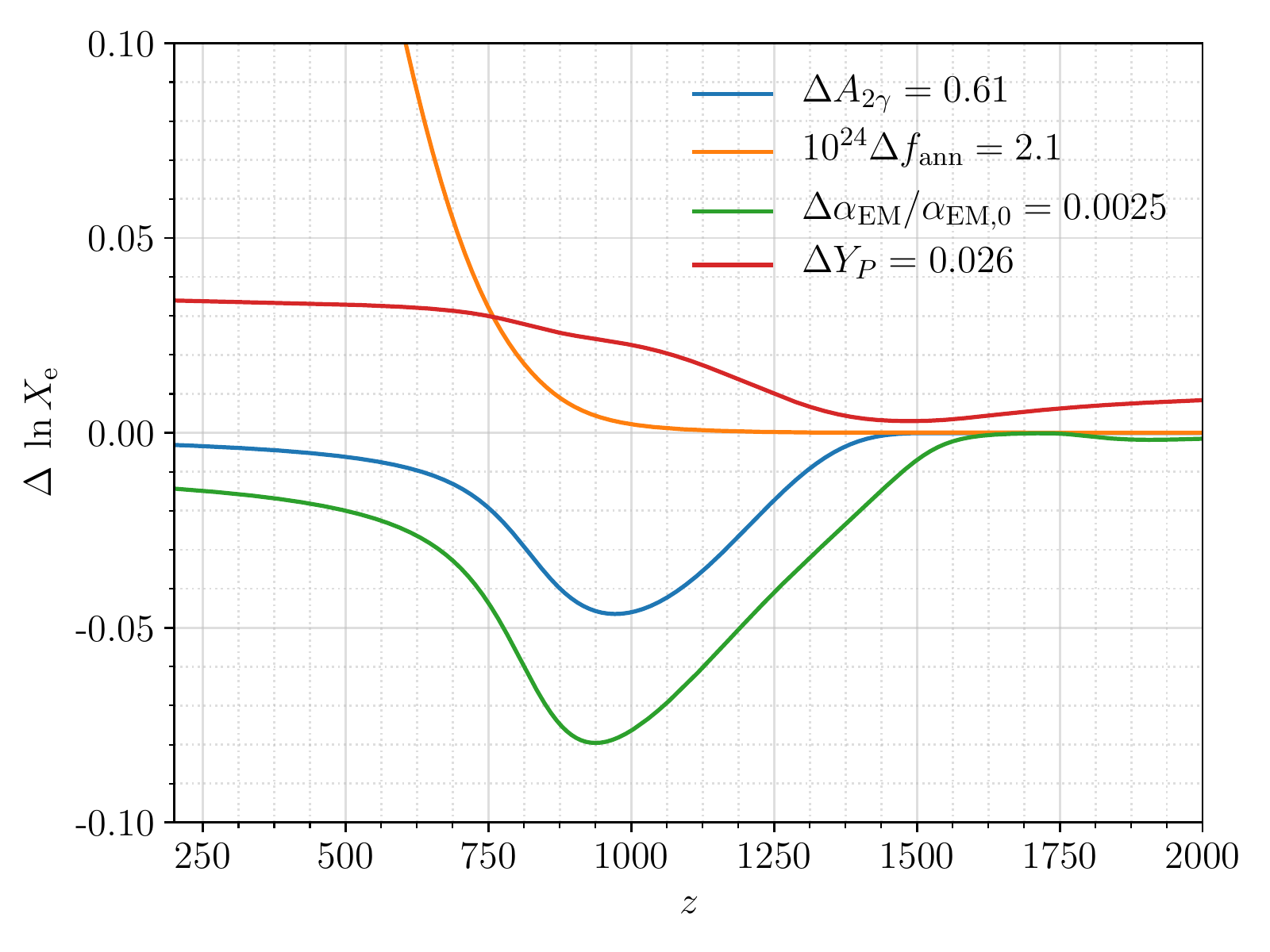}
  \caption{Relative changes in the electron fraction $\xe$ due to same physical parameter changes shown in Fig.~\ref{fig:vertical_projections}. The variations have been scaled by the standard deviations constrained with \planck. The curves were obtained using {\tt CosmoRec}.}
  \label{fig:projections}
\end{figure}

To illustrate matters, in Section~\ref{sec:proj_results} we will examine several single-parameter extensions to the standard recombination model: the effective decay rate for the hydrogen ${\rm 2s\rightarrow1s}$ transition during recombination, $\Atwo$; the primordial abundance fraction of helium, $\yp$; the dark matter annihilation efficiency parameter, $\fann$; and the fine structure constant, $\alpha_{\rm EM}$. 
The corresponding normalised projections, $\rho_i$, for various cases are summarised in Table~\ref{tab:projections}. All the shown step sizes, $\Delta\parA$, yield numerically stable responses in the linear regime of $\dxe$ shown in Fig.~\ref{fig:projections}.

 \begin{table}
     \centering
     \begin{tabular}{l|c|c|c|c}
        \hline\hline
        Parameter ($\mathcal{A}$) & $\Delta\parA$ & $\hat{\rho}_1$ & $\hat{\rho}_2$ & $\hat{\rho}_3$ \\
        \hline
        $\Atwo$ & $0.04$ & $0.16$ & $0.15$ & $0.05$ \\[0.5mm]
        $\yp$ & 0.004 & $-1.67$ & $0.83$ & $-8.64$ \\[0.5mm]
       $10^{24}\fann$ & $0.2$ & $7.47\times10^{-5}$ & $0.0040$ & $-0.0095$ \\[0.5mm]
        $\aEM$ & $0.0002$ & $35.42$ & $10.62$ & $63.90$ \\
        \hline
     \end{tabular}
     \caption{Projections $\hat{\rho}_i$ of physical $\dxe$ changes onto the \planck eigenmodes alongside the parameter step size $\Delta\parA$ used. 
     Each value $\hat{\rho}_i$ measures how strongly the physical variations in Fig.~\ref{fig:projections} projects onto our \planck modes in Fig.~\ref{fig:planck}}
     \label{tab:projections}
 \end{table}
The two-photon decay rate of hydrogen controls one of the key recombination channels through which the Universe becomes neutral \citep{Zeldovich68, Peebles68}. Increasing the two-photon decay rate caused an acceleration of the recombination process (see Fig.~\ref{fig:projections}) and a positive eigenspectrum $\hat{\rho}_i$ with the largest projection onto PC 1 (see Table~\ref{tab:projections}). Comparing the values of $\hat{\rho}_i$ to the errors on $\mu_i$, we can anticipate that the first two modes will drive the PCA constraint.

Changing $\yp$ has the main effect of modifying the ratio of hydrogen to helium atoms. Since $\xe\simeq (1-\yp/2)/(1-\yp)$, we have $\dxe\simeq (1/2)\Delta\yp$ with additional corrections. Thus, increasing $\yp$ leads to a $\dxe>0$ at all redshifts (see Fig.~\ref{fig:projections}), with the largest projection on PC3. In this case, we can anticipate the modes PC1 and PC3 to drive the direct projection.

Dark matter annihilation results in a delay of recombination \citep[e.g.,][]{Chen2004, Padmanabhan2005}, which is particularly noticeable at low redshifts (see Fig.~\ref{fig:projections}). 
We use the $\fann$ parametrisation defined in previous recombination papers\footnote{This relates to the annihilation parameter, $p_{\rm ann}$, used in the \planck parameter paper \citep{Planck2015params} by
$\fann\approx 8.2\times10^3p_{\rm ann}$. However, further differences in the energy deposition functions and the on-the-spot approximation complicate the direct comparison \citep[e.g.,][]{Slatyer2009, Galli2013}.} \citep{Chluba2010a}. 
As expected, the projection onto PC1 is much smaller in comparison to the others, and consequently we expect weak constraints from the first mode alone.

Finally, $\aEM>1$, where $\alpha_{\rm EM,0}$ denotes the local value of the fine-structure constant, causes an acceleration of recombination \citep[e.g.,][]{Kaplinghat1999, Hart2017} with significant projections onto PC1 and PC3. Bigger improvement in the PCA projection constraint are thus expected from PC3 rather than PC2.

\vspace{-3mm}
\subsection{Optimising the projections for parameter constraints}\label{sec:proj_opt}
 Using the definitions of the projections and the renormalised projection $\hat{\rho}_i$, we can use our $\chi$-squared definition in Eq.~\eqref{eq:chi2} to minimise the parameter change, $\delA$ as a distance from the reference parameter change $\delA_0$. Redefining and substituting Eq.~\eqref{eq:rho_new} into Eq.~\eqref{eq:chi2}, we obtain 
  \begin{equation}\label{eq:chi2_new}
    \chi^2=\left(\delA\,\hat{\rho}_i-\mu_i\right)^{\rm T}\Sigma_{ij}^{-1}\left(\delA\,\hat{\rho}_j-\mu_j\right).
 \end{equation} 
Given that this describes a simple Gaussian `likelihood', the $\chi$-squared is parabolic and therefore, if we find the minima, we can project the eigenmode amplitudes onto a parametrised amplitude change in $\parA$. Taking the first derivative of Eq.~\eqref{eq:chi2_new} with respect to the model parameter yields
  \begin{align}
  \label{eq:chi2_deriv}
       \partial_\Delta\chi^2 &= \frac{\partial}{\partial \delA}\left[\left(\delA\,\hat{\rho}_i-\mu_i \right)^{\rm T}\Sigma_{ij}^{-1}\left(\delA\,\hat{\rho}_j-\mu_j\right)\right], \nonumber\\
    &= 2\hat{\rho}_i\Sigma_{ij}^{-1}\left(\delA\,\hat{\rho}_j-\mu_j\right) \equiv 0.
 \end{align}
As a result, the best fit variation of the parameter, $\delA_{\rm bf}$ can be used to estimate the best-fit value, $\parA_{\rm bf} = \parA_{\rm fid}+\delA_{\rm bf}$.
 
We can also use the second derivative of the $\chi$-squared to calculate the error on the given parameter. This is from assuming that the $\chi$-squared can be related to a Fisher matrix and subsequently, a standard deviation by the usual method (see Kramer-Rao bound for more details). Using $\chi^2 = -2\ln\mathcal{L}$, we have
  \begin{align}
     F_{\parA} =\frac{1}{2}\partial^2_\Delta\chi^2 &= \frac{1}{2} \frac{\partial}{\partial \Delta\parA}\left[2\hat{\rho}_i\Sigma_{ij}^{-1}\left(\Delta\parA\,\hat{\rho}_j-\mu_j\right)\right], \nonumber \\ 
     &= \hat{\rho}_i\,\Sigma_{ij}^{-1}\hat{\rho}_j,
\end{align}
leading to an error estimate on the parameter $\parA$, 
\begin{align}\label{eq:projerr}
    \sigma_\parA &\simeq \sqrt{ F_{\parA}^{-1}}
    \simeq \sqrt{(\hat{\rho}_i\,\Sigma_{ij}^{-1}\hat{\rho}_j)^{-1}}.
\end{align}
The projection method offers us the power to constrain physical models that vary functions such as the ionization history, $\xe$ with a single set of constrained eigenmode amplitudes $\mu_i$ for a given dataset. The constraint requires a simple linear algebra calculation as described in \Cref{eq:chi2_deriv,eq:projerr}.
For close to diagonal mode covariance, one then simply has
$\sigma_\parA \simeq (\sum \hat{\rho}^2_i/\sigma(\mu_1)^2)^{-1/2}$, which for our modes yields nearly identical results.

\vspace{-3mm}
\subsection{Using the direct projections with \planck modes}\label{sec:proj_results}
 We are now in the position to use the projection method outlined in Sec.~\ref{sec:proj_def} to recreate direct physical parameter constraints for the considered models. This method is expected to work best for physical scenarios that primarily affect the ionization history $\xe$ (e.g., the two-photon decay rate), but otherwise have no effect on the CMB anisotropies. We furthermore expect mechanisms that mainly affect the recombination history around $z\simeq 1100$ to be best-constrained, as the first few modes mainly pick up information from that range.
 
  \begin{table}
     \centering
     \begin{tabular}{l c c c c}
     \hline\hline
         Parameter ($\parA$) & MCMC result & 3 mode projections
         \\
         \hline
          $\Atwo$ & $7.72\pm0.60$ & $7.60\pm0.64$ \\[0.5mm]
          $\yp$ & $0.250^{+0.026}_{-0.027}$ & $0.277\pm0.035$ \\[0.5mm]
          $10^{24} \fann$ ($95\%$ CL) & $<4.3$ & $<6.2$ \\[0.5mm]
          $\aEM$ & $0.9988\pm0.0033$ & $0.9969\pm0.0029$ \\
          \hline
         \hline
     \end{tabular}
     \caption{The parameter constraints on $\parA$ for the full MCMC result for \planck 2015 TTTEEE + lowTEB compared with the projections result using the first three eigenmodes from the given analysis. The variations of the given $\parA$ in the ionization history are shown in Fig.~\ref{fig:projections}. The \planck constraint for $\aEM$ is only from $\xe$ effects, neglecting effect from rescaling of the Thomson cross section (see text for more details).}
     \label{tab:projection_results}
 \end{table}

\begin{figure*}
     \centering
     \includegraphics[width=0.9\linewidth]{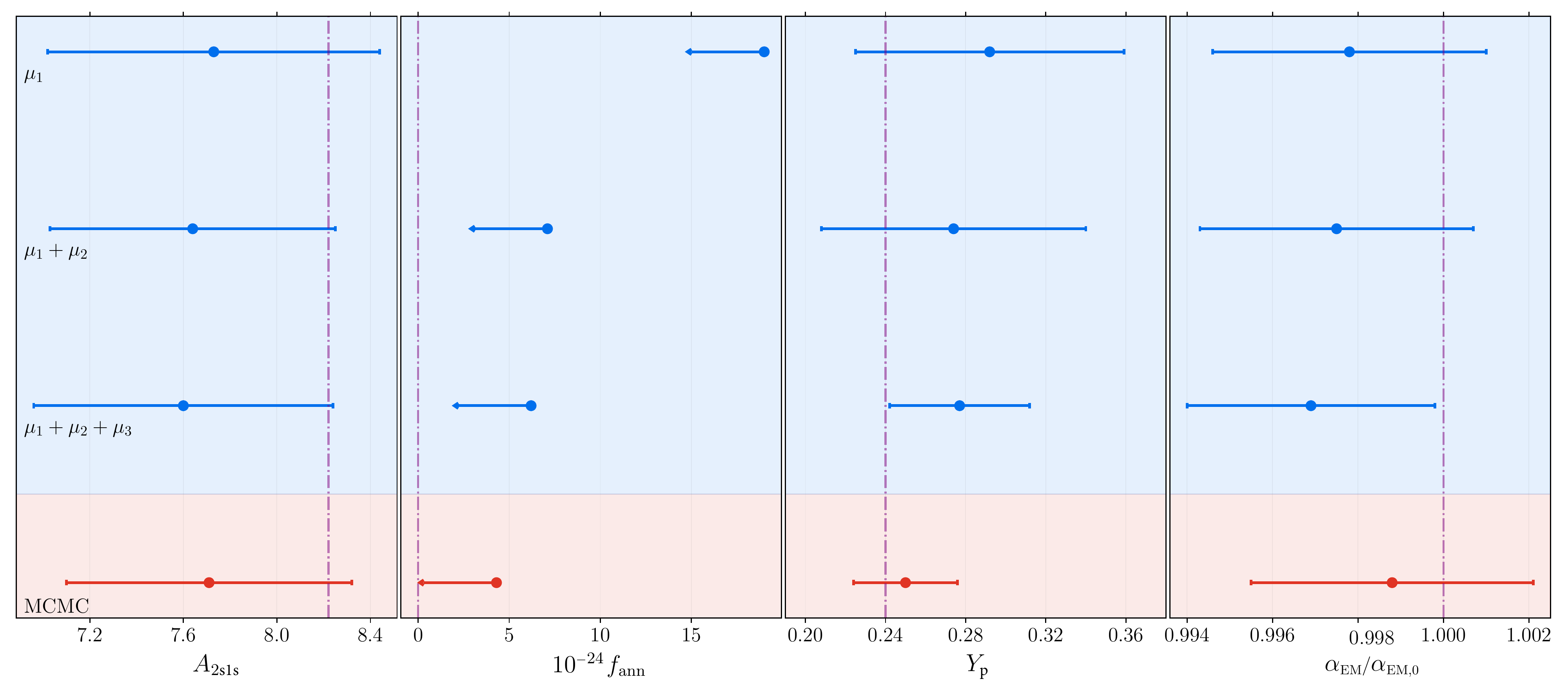}
     \caption{The direct projection method explained in Sec.~\ref{sec:proj_def} produces these results. Here we show the best error constraints for 4 commonly constrained parameters: two photon decay rate ($\Atwo$), primordial helium fraction ($\yp$), dark matter annihilation efficiency ($\fann$) and fine structure constant ($\aEM$). The top results ({\it blue}) are the best projections made with the \planck constrained eigenmodes. The bottom result ({\it red}) is the MCMC result gained from many previous papers \citep{Planck2015params,Hart2017}. Here we show the $95\%$ upper limit for $\fann$ as explained in Table~\ref{tab:projection_results}.} 
     \label{fig:vertical_projections}
 \end{figure*}

The projection results from applying the first three principal components are summarized in Table~\ref{tab:projection_results}. We also give the values from the direct MCMC runs for \planck 2015 TTTEEE + lowTEB for reference. To further illustrate how the method works, we show the evolution of the projection values and errors when varying the number of modes in Fig.~\ref{fig:vertical_projections}. 

The comparison between our projections and the MCMC with \planck data shows that the obtained error estimates and central values are very consistent (Table~\ref{tab:projection_results}). For $\Atwo$, the projection error is only $\simeq 0.06\,\sigma$ away from the MCMC result and also the central value is well within the error. The first and second mode are driving the constraint, while addition of the third mode has a minor effect (Fig.~\ref{fig:vertical_projections}). This is expected from the fact that $\hat{\rho}_3/\sigma(\mu_3)\simeq 0.14$ while for the other modes we find $\hat{\rho}_1/\sigma(\mu_1)\simeq \hat{\rho}_2/\sigma(\mu_2)\simeq 1$.

For variations of $\aEM$, the projection error and central value are again extremely close to the MCMC equivalent (error agrees to $\simeq0.1\sigma$). However, here it is important to mention that we neglected the effect of $\alpha_{\rm EM}$ on the Thomson cross section, which has a $\simeq 30\%$ effect on the error \citep{Hart2017}. Including this effect would affect the CMB anisotropies in a way that is independent of $\xe$ and thus needs to be modeled separately. \test{This effect will be modeled on the upcoming paper where we apply this technique to fundamental constants.}

The deviation of the PCA projection results from the MCMC run is larger for $\yp$. The PCA error is about $\simeq 40\%$ larger and the central value of $\yp$ (although consistent to within the errors) lies above the MCMC result (see Fig.~\ref{fig:vertical_projections}).
We can explain this by considering Fig.~\ref{fig:projections}. The variations arising from $\yp$ are most visible at low and high redshifts, while the first three modes are mainly localised around $z\simeq 1100$. 
Though some information at the extreme redshifts is contained in the first three eigenmodes, non-negligible projections are expected onto higher modes. 
This loss of information into the higher modes is likely the cause of the departures.
This statement is further supported by the fact that the third mode is indeed strongly driving the obtained PCA constraint (see Fig.~\ref{fig:vertical_projections}).
The addition of the fourth PC could thus likely improve the PCA constraint, but was not considered here.

The story is very similar for the annihilation scenario. The final PCA constraint is slightly weaker than the MCMC result, showing that more information from the tails of $\dxe$ is contained in the modes beyond the third.
As anticipated in Sect.~\ref{sec:rho_i_results}, PC1 has little constraining power, while adding PC2 gives a large improvement (see Fig.~\ref{fig:vertical_projections}).

We shall mention a few details about using the PCA method to derive model parameter constraints. 
We found the final projection results to be mostly independent of the precise background cosmology used in generating the $\Delta\ln\xe$ projections to the $1\%$ level. Fully including the drift of the standard parameter values in the $\chi^2$ calculation also had a small impact on the final constraint.
Finally, we stress again that for a parameter to be effectively constrained using the projections method, the physical change that is studied has to mainly affect the projection variable. In this case, we are projecting onto recombination histories ($\Delta\ln\xe$) and the physical models shown in Fig.~\ref{tab:projections} all only affect recombination. \test{In the case of $\aEM$, this also affects the Thomson cross section; however, this can be accounted for through an effective rescaling of $\xe$ \citep[see][]{Hart2017}}. However, if the variation affects the CMB anisotropies outside of the recombination calculation, the result is only going to be partially correct. For example, changes in the average CMB temperature $\TCMB$ shift the redshift of recombination which has a clear effect on recombination. However, there are other effects such as the amplitude of the CMB power spectrum, the photon energy density and the matter-radiation equality. These effects will not be captured by the $\xe$ projections alone and require more detailed modeling.

Overall, we find the PCA projection method works as a good and efficient estimation tool. 
It allows us to quantify the projections of our eigenmodes with real, physical variations during recombination. This is an exciting way to determine which components are most strongly correlated with non-standard recombination physics and estimate the parameters encompassing these variations in a simple way. \test{It furthermore allows us to efficiently obtain parameter constraints together with CMB spectral distortion \citep[e.g.,][for overview]{Chluba2019WP}, for which detailed  case-by-case computations (i.e., for photon and particle injection scenarios) can have runtimes that are prohibitive for full MCMCs \citep[e.g.,][for numerical details]{Chluba2011therm}.}

\section{Conclusion}
We presented an improved principal component analysis of modifications to the standard recombination history, introducing the new code package {\tt FEARec++}. 
{\tt FEARec++} allows a careful stability analysis for the numerical derivatives required for the creation of the eigenmodes. It automatically generates high-quality principal components using real data from \planck, ensuring these are orthogonal to high precision (better than $\simeq 0.1\%$) even after parameter marginalisation. 
Applying these eigenmodes to \planck 2015 data, we confirm that standard recombination is statistically favoured (see Table~\ref{tab:planck}). 
This is in agreement with previous analysis \citep{Planck2015params, Planck2018params}, which is based on the method developed by \citet{Farhang2011, Farhang2013}.

The main improvement of {\tt FEARec++} over previous work is a more direct inclusion of the \planck likelihood in the construction of the modes. Several numerical obstacles (see Sect.~\ref{sec:formalism} for more details) had to be overcome to ensure that high-quality modes are obtained. This in particular affected the third eigenmode and the residual correlations among the modes, which remain below the $\simeq 0.01\%$ level right after construction (see Table~\ref{fig:contours_degen}).
The improved standard parameter decorrelation yields a reduced error ($\simeq 2.5$ times) for the third principal component in comparison with the previous analysis \citep{Planck2015params}.
Furthermore, our recombination eigenmodes have shown that adding datasets such as CMB lensing and baryonic acoustic oscillations from SDSS do not affect the mode constraints significantly (see Table~\ref{tab:planck_data}). 

We also presented a direct projection method (Sect.~\ref{sec:proj}), which convincingly replicates the MCMC limits using our mode constraints. It also allows us to specifically see which eigenmodes contribute most to each physical variation (shown in Fig.~\ref{fig:vertical_projections} and Table~\ref{tab:projections}).
This method can be used to obtain reliable estimates for physical models that directly affect the cosmological recombination history, avoiding to have to rerun the full MCMC. We attempted to implement the projections method with the older modes; however non-orthogonalities between the eigenmodes and larger errors in the third eigenmode led to inconsistent results. A module to calculate these estimates was added to {\tt FEARec++} to quickly explored the allowed parameter space. It should provide useful alternative to the full \planck likelihood when studying non-standard recombination scenarios related to decaying and annihlating particles together with future limits on CMB spectral distortions \citep[e.g.,][]{Chluba2011therm, Chluba2013PCA}.

Albeit all the successes, there are a few open issues. 
So far we have been unable to go beyond the third recombination eigenmode. The main limitation is the numerical precision of the likelihood code (see discussion in Appendix~\ref{app:optimise}), which prevented us from obtaining a fully decorrelated mode. This may be remedied by the \planck 2018 likelihood code, which was made public this past summer. However, we have not explored this yet.
We furthermore still had difficulties when optimising the modes for \planck with external datasets. In these cases, the third mode was found to degrade. However, since the \planck-only modes performed extremely well on all data combinations considered here (i.e., showing no significant degradation of the errors or spurious correlations), 
we leave it to future work to stabilise the eigenmodes when more data is included in the construction. 

Given the new methods implemented for {\tt FEARec++}, we plan to extend our analysis to time-dependent variations of $\aEM$ and $\me$. In \citet{Hart2017} we showed that \planck data is indeed sensitive to these effects. Given recent considerations regarding a possible connection of variations in $\me$ to the Hubble tension \citep{Hart2019}, this idea seems timely.
This will also be a useful test for the 2018 \planck likelihood.  
Finally, we are now in the position to combine recombination and reionization histories into a dual code, which can solve the full free electron fraction. A full analysis that encompasses both low- and high-redshift information in the free electron fraction will allow us to apply the projection method to physical parameters surrounding reionization. \test{The direct projections will also be an excellent tool for parameter estimations combined with CMB spectral distortions \citep[e.g.,][for recent overview]{Chluba2019WP}, for which case-by-case computations can still be time-consuming.}

\vspace{-3mm}
\small
{\section{Acknowledgements}
We would like to cordially thank Marzieh Farhang for providing the \planck 2015 modes as well as discussions surrounding the original methodology and how the modes were constructed for the PCA. 
We are also grateful to Silvia Galli, Antony Lewis and Eleanora Di Valentino for discussions surrounding the stability of the \planck likelihood and the general use of the likelihood code {\tt CosmoMC}.
We extend thanks to Richard Rollins for useful discussions when including the {\tt Eigen} library for the eigenanalysis and to Thejs \,Brinckmann for very useful discussions surrounding the likelihood\, function as a mechanism for a Fisher matrix. 
\test{We also thank the reviewer for their feedback on the clarity of subtleties within the PCA methodology.} 
We would also like to thank  the Jodrell Bank Centre for Astrophysics for the use of the {\sc Fornax} cluster, and Anthony Holloway for his support when using this resource. 
LH is funded by the Royal Society through grant RG140523.
JC is supported by the Royal Society as a Royal Society University Research Fellow at the University of Manchester, UK.
This work was also supported by the ERC Consolidator Grant {\it CMBSPEC} (No.~725456) as part of the European Union's Horizon 2020 research and innovation program.
}

\vspace{-3mm}
\small 
\bibliographystyle{mn2e}
\bibliography{Lit}

\vspace{-3mm}
\appendix
\section{Generating orthonormal basis functions and the responses}
\label{app:basis}
In order to see which variations the CMB data favours most we need to create an orthonormal set of response functions. These functions form their own vector space and these shall map from perturbations in the recombination history to responses in the CMB power spectrum. For the eigenmodes to be orthogonal, the basis functions must form an orthonormal set as well. Previous works show we can choose many different natural functions to form the eigenmodes \citep{Farhang2011}. Therefore, we shall use Gaussian shapes since they are smooth and easy to implement into the Boltzmann equation solver. As described in \eqref{eq:basis},
\begin{equation}
  \left(X_{\rm e}\right)_i = X_{\rm e}\left[1+\varphi_i(z)\right],
\end{equation}
where $\varphi_i(z)$ is given by,
\begin{equation}
    \varphi_i(z) = \frac{1}{\sqrt{2\pi \sigma^2}}\exp\left[-\frac{1}{2}\left(\frac{z-z_i}{\sigma}\right)\right]^2.
\end{equation}
The setup that generates the adjacent basis functions have left large overlaps in the functions in previous papers. We have attempted to generate eigenmodes with basis functions separated at two different levels. Firstly, following the previous work \citep{Farhang2011} we have formed basis functions with widths defined by,
\begin{equation}
    w_1 = \frac{z_N-z_1}{2N_{\rm func}\sqrt{2\ln2}} = \frac{\delta z}{2\sqrt{2\ln2}},
\end{equation}
where $z_N$ and $z_1$ are the limits of our redshift range and $\delta z $ is the spacing between the peaks. 
Secondly, we used a thinner width $w_3 = w_1/3$. The eigenmodes were not affected by the thinner width of the basis functions but we did not assess the mode creation for larger widths where more than 2 functions overlap. 
The basis functions should cover the full redshift space however since we are looking at phenomena during recombination, we can localise these functions to a redshift range $z = \{300,2500\}$. Outside this range, the CMB anisotropies are changed negligibly unless we enter the reionsation era at $z\lesssim 10$.
The basis functions are propagated through the recombination history via {\tt CosmoRec} and then applied to the Boltzmann code {\tt CAMB}. In this analysis we analyse the responses from the CMB temperature and polarisation power spectra (see Fig.~\ref{fig:cvl} for examples). 

\begin{figure}
  \centering
  \includegraphics[width=\linewidth, trim={0 5 0 0},clip]{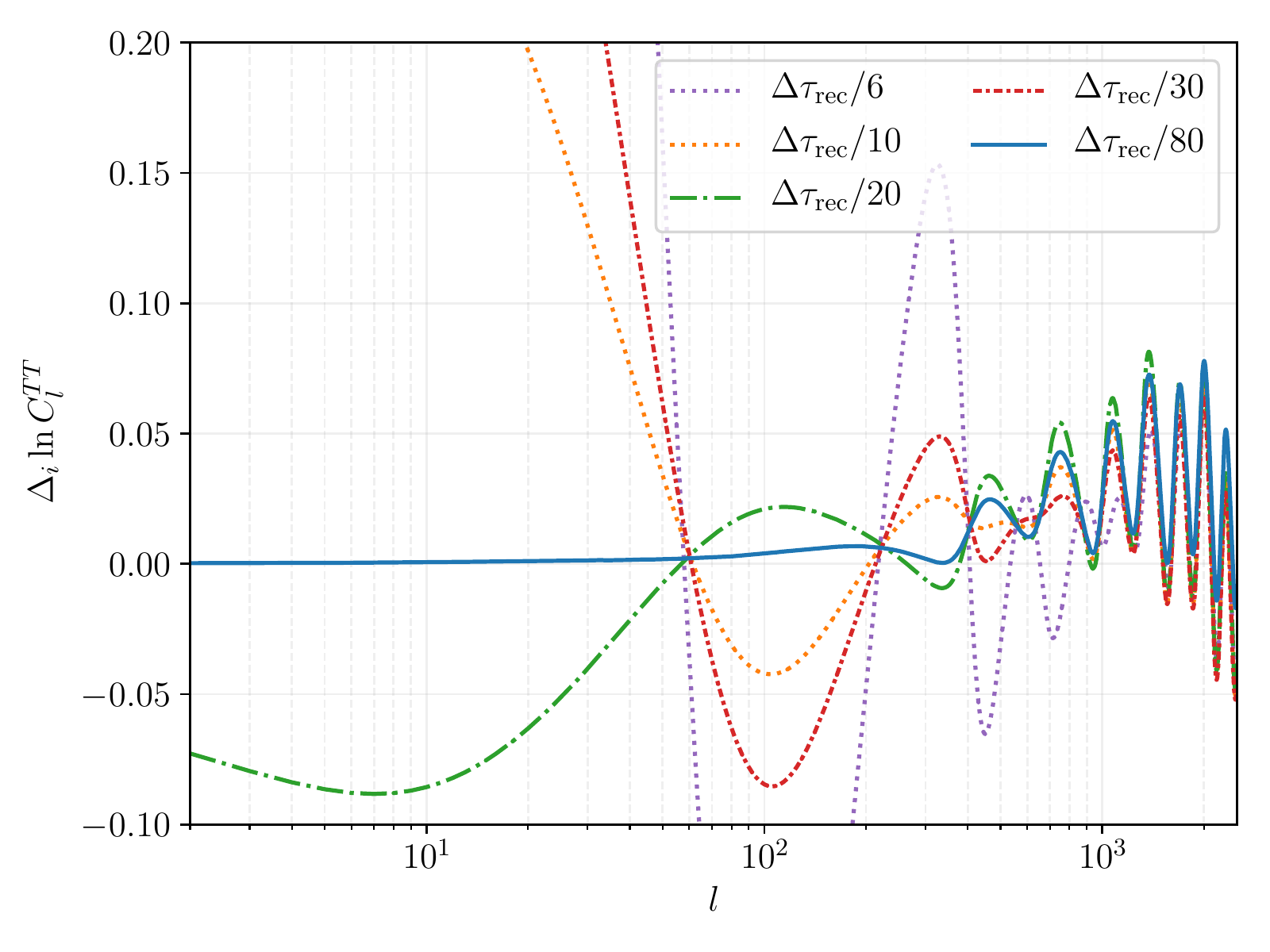}
  \caption{The normalised responses from the CMB power spectra from a basis function pivoted around $z_i = 1100$ for different accuracy levels of the time sampling during recombination. As $\Delta\tau_{\rm rec}$ becomes smaller, the resolution during recombination becomes finer.}
  \label{fig:dtau}
\end{figure}

\begin{figure}
  \centering
  \includegraphics[width=\linewidth, trim={0 5 0 0},clip]{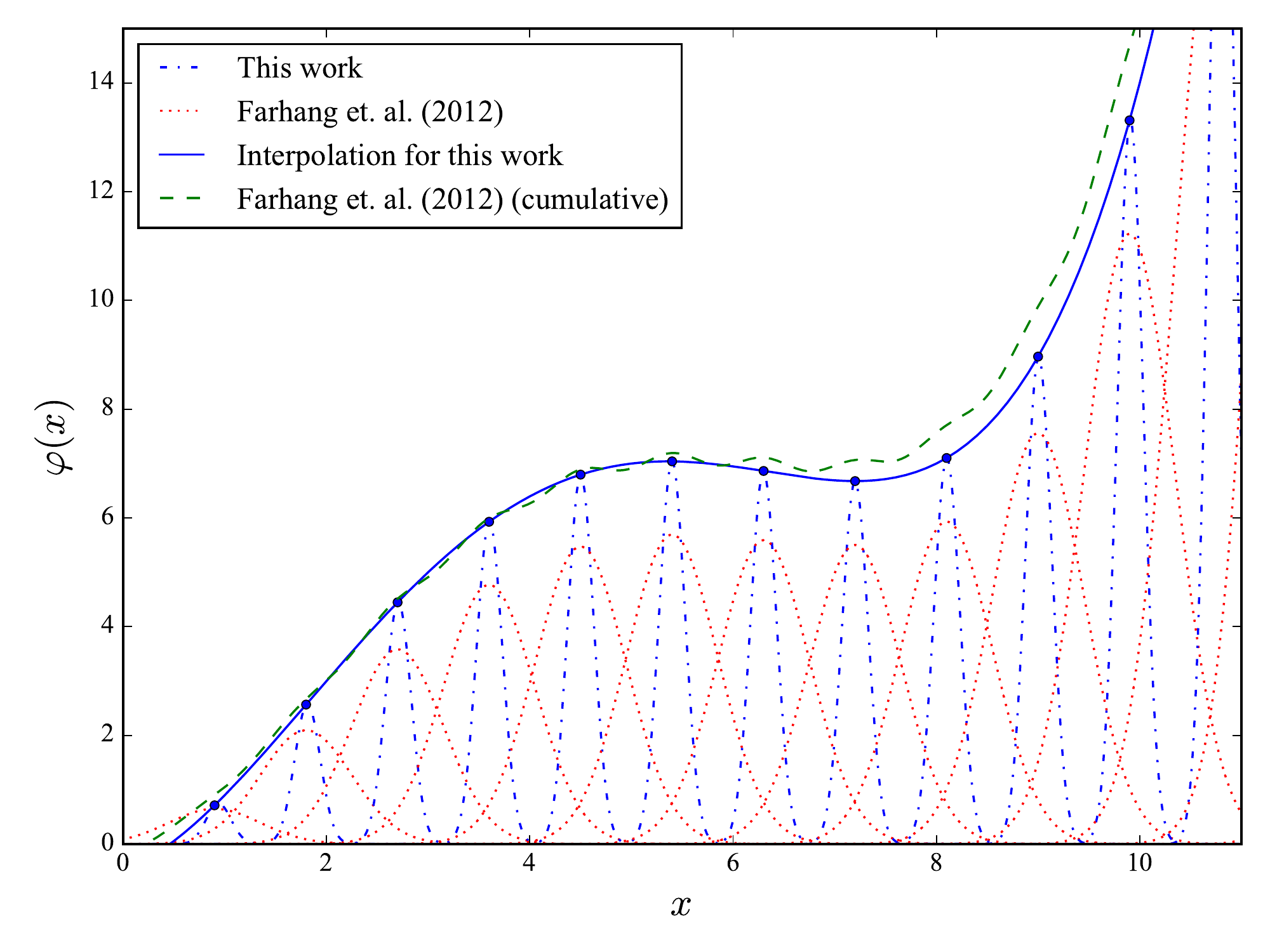}
  \caption{The different basis conditions for this work and \citealt{Farhang2011}. The interpolation between Gaussian peaks for this work (blue, solid) and the cumulative Gaussians approach from previously (green, dashed) are also shown. Departures from the input function in the latter approach are largest where the function is steep.}
  \label{fig:basis}
\end{figure}

\vspace{-3mm}
\subsection{Numerical response of the Boltzmann code} 
\label{app:dtau}
In this section, we present a numerical study of the stability of CMB anistropy responses. The derivatives have to be numerically stable to accurately trace out the CMB responses from different redshifts. 
In Fig.~\ref{fig:dtau}, we show how the timestep during recombination, $\Delta\tau_{\rm rec}$, has an explicit effect on the derivatives found. This is due to the resolution around recombination not being fine enough to sample the basis functions, leading to discontinuities in the Boltzmann equations solver. The numerical instabilities arise from not only the sampling of the recombination history $\xe$ but also the slope $\partial\xe/\partial z$, which affects the Thomson visibility function.  When we rescale this timestep to a finer resolution, the derivatives converge and we obtain results consistent with the previous work \citep{Farhang2011}. Once these instabilities are fixed, we obtain the $C_l$-responses shown in Fig.~\ref{fig:responses}. 
Flexibility in changing the $\Delta\tau_{\rm rec}$ parameter is now explicitly available in {\tt CAMB} and the Python wrapper {\tt PyCamb}, so this can be easily implemented in the future.

\vspace{-3mm}
\section{Analytical method for creating fisher matrices}
\label{app:analytic}
The optimised responses from the CMB angular power spectra can be used to create a Fisher matrix. This matrix contains the effective signal-to-noise for each of the responses from all the different redshifts $z_i$. The Fisher matrix is formulated using the vector of CMB power spectra derivatives, $\partial_i C_l^{X}$, and the modelled CVL covariance matrix, $\Sigma$, as,
   \begin{equation}
     F_{ij} = \sum_{l=0}^{l_{\rm max}} \partial_i C_l\cdot \Sigma^{-1}_l\cdot\partial_j C_l\,. 
   \end{equation}
The vectors of responses from the CMB temperature, polarisation and cross-correlated power spectra respectively are,
   \begin{equation}
     \partial_i C_l = 
     \begin{pmatrix}
       \partial_i C_l^{\rm TT} \\ \partial_i C_l^{\rm EE} \\ \partial_i C_l^{\rm TE} \\ 
     \end{pmatrix},
     \label{eq:diff}
   \end{equation}
whilst the covariance matrix for a given $l$, $\Sigma_l$ is, 
   \begin{equation}
  {\Sigma_l} = \frac{2}{2l+1}
  \begin{pmatrix}
    C^{\rm TT^2} & C^{\rm TE^2} & C^{\rm TT}C^{\rm TE} \\
    C^{\rm TE^2} & C^{\rm EE^2} & C^{\rm TE} C^{\rm EE} \\
    C^{\rm TT}C^{\rm TE} & C^{\rm TE}C^{\rm EE} & \frac{1}{2}\left(C^{\rm TE^2}+C^{\rm TT}C^{\rm EE}\right) \\
  \end{pmatrix}.
  \label{eq:matrix}
\end{equation}
This formulation of the Fisher matrix with two signal vectors ($\partial_i C_l$) and a noise covariance matrix ($\Sigma_l^{-1}$) has previously been shown in \citet{Verde2009}. It has also been used in previous principal component analyses with the CMB \citep{Finkbeiner2012}


Diagonalising this matrix $F_{ij}$ will give us a set of eigenvectors that, when recast onto our basis functions, define the most constraining variations of $\xe$ given a dataset from the CMB (in this case, an analytical noise simulation for a CVL experiment). 

\subsection{Interpolating through the diagonalised basis functions}
\label{app:interpolation}
To allow the principal components to smoothly modify the recombination history $\xe$, we need to make the discrete eigenvectors continuous. 
%
A set of basis functions that define a generic curve are shown in Fig.~\ref{fig:basis}. We show the basis functions used in this paper and also the basis functions that were used in \cite{Farhang2011}. In that work, the functions were summed and the cumulative function was smoothed using a Gaussian-filter. In this work, we take the rescaled peaks of the basis functions from the eigensolver and then smoothly interpolate between them using a cubic spline interpolation routine. 
This approach gives us extremely orthogonal eigenmodes, where the alternative routine gives us spurious correlations and also does not accurately recreate the eigenmodes where the gradient is large (see Fig.~\ref{fig:basis}). All the principal components presented throughout this paper use this interpolation approach.

\begin{figure}
  \centering
  \includegraphics[width=0.9\linewidth, trim={0 40 0 0},clip]{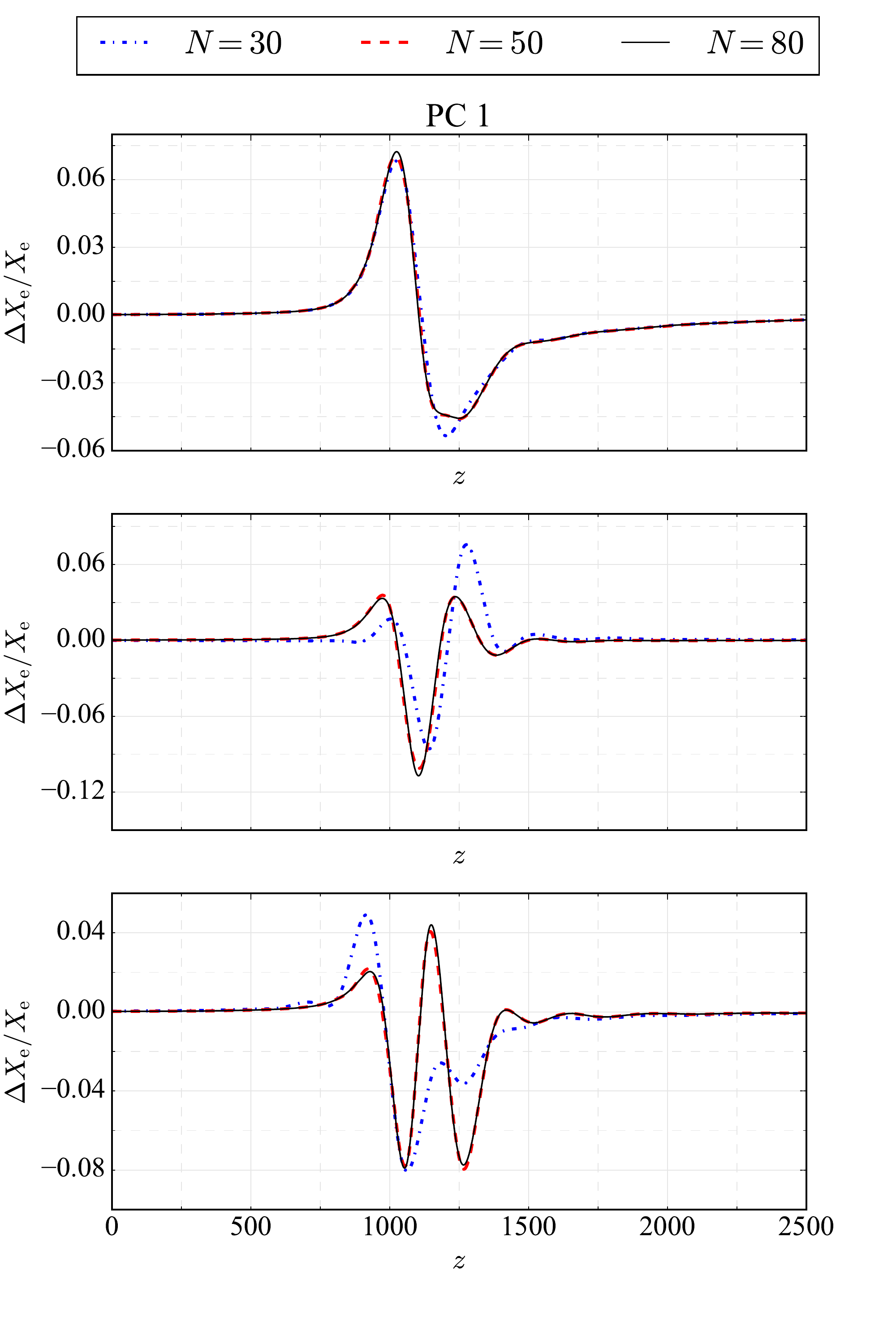}
  \caption{The first three $\xe$ CVL principal components in redshift space for various numbers of basis functions, $N=\{30,50,80\}$ across the redshift range $z_i = \{300,2500\}$.}
  \label{fig:number_test_cvl}
\end{figure}

\subsection{Eigenmode resolution}
\label{app:resolution}
Another important aspect to consider is the resolution of the mode space we create, as described in Appendix \ref{app:basis}. Not only must a reasonable bound of parameter space (in this case in redshift $z$) be chosen, but the number of functions that form the basis set must be large enough that we capture the full shapes of the principal components. 

In Fig.~\ref{fig:number_test_cvl},  the eigenmodes show little change once the number of basis functions for our treatment exceeds $N\simeq 50$. It is shown in Fig.~\ref{fig:number_test_cvl} that non-negligible variations lie in the region $z<2500$.  Here we use $N=80$ to be consistent with the approach used in the most recent \planck publication \citep{Planck2018params}. 
For these modes, we have used the same range of redshifts, $z_i=\{300,2500\}$, to cover the full redshift space that is affected by the recombination processes important to the production of the CMB anisotropies.

\section{Direct likelihood method for creating Fisher matrices}
\label{app:direct}
The likelihood function across all variables can be used to generate principal components as an alternative to the analytic method. We coin the term `{\it direct method}' to describe this as we are directly sourcing the likelihood function to test the data's response to the change in model. This is taken from the definition of the Fisher matrix $F_{ij}$ in terms of the likelihood function $\mathcal{L}$ (see \Cref{eq:fij_definition}).

The code from \planck can give us the $\DelL$ values directly or it can be calculated through the MCMC package {\tt CosmoMC} \cite{COSMOMC, Planck2015like}. This will enable us to calculate the observed Fisher information matrix.\footnote{Note in the analytical case, we've calculated the expected Fisher information matrix.} The three-point stencils are well defined however for reasons outlined below in Appendix~\ref{app:stability}, we need higher order derivatives. Here we use the higher order stencil,
\begin{equation}
  F_{ii} = \frac{\partial^2 L}{\partial q_i^2} = \frac{\left(-L_2+16L_1 - 30L_0+16L_{-1}-L_{-2}\right)}{12\,\delta q_i^2},
\label{eq:Fii}
\end{equation}
for the diagonal elements of the Fisher matrix where $L = \ln\mathcal{L}$, the log-likelihood function, and the notation $L_k = L(p_0+k\delta p)$ defines the function at a given step away from the fiducial case. The higher order stencil for the off-diagonal elements is,
\begin{align}
\begin{split}
  F_{i\neq j} &= \frac{\partial^2 L}{\partial q_i \partial q_j} \\
  &= \frac{1}{144\,\delta q_i\delta q_j} \sum_{k=-2}^2\sum_{m=-2}^{2}\left(-1\right)^{(k+m)}\frac{km}{|k||m|}\,2^{3(4-|k|-|m|)}\, \Lvec{k}{m}
  \end{split}
  \label{eq:Fij}
\end{align}
where $\Lvec{k}{m} = \ln\mathcal{L}\left(p_0+k\delta p, q_0+m\delta q\right)$ is the given deviation of the likelihood function according to parameter changes $\delta p_i$ and $\delta p_j$. One should bear in mind that the stencil in \Cref{eq:Fij} requires 16 numerical evaluations of the likelihood. Reducing the runtime of this is key and one of the biggest breakthroughs with {\tt FEARec++} (see Sec.~\ref{sec:fearec}).

\subsection{Troubleshooting the stability of \LCDM parameters within the likelihood}
\label{app:stability}
\begin{figure}
\centering
\includegraphics[width=\linewidth, trim={0 30 0 0},clip]{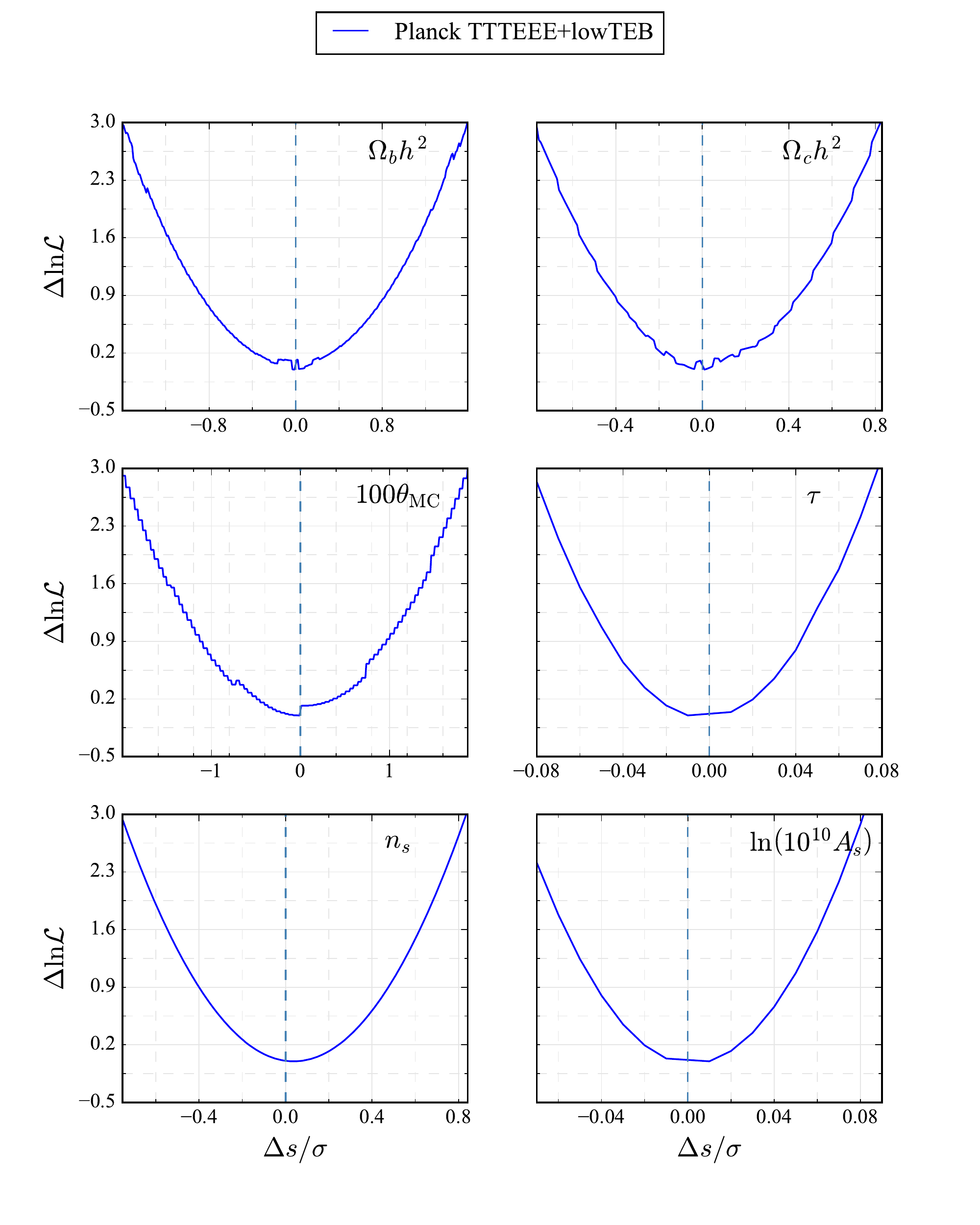}
\caption{Differences in the likelihood values close to the minima of the fiducial point. The minima of the likelihood for each parameter (blue dashed) has been added. The parameter curves have all been scaled such that the $\Delta\ln\mathcal{L}\simeq3$.}
\label{fig:stability_curve}
\end{figure}

\begin{figure}
\centering
\includegraphics[width=\linewidth, trim={0 30 0 0},clip]{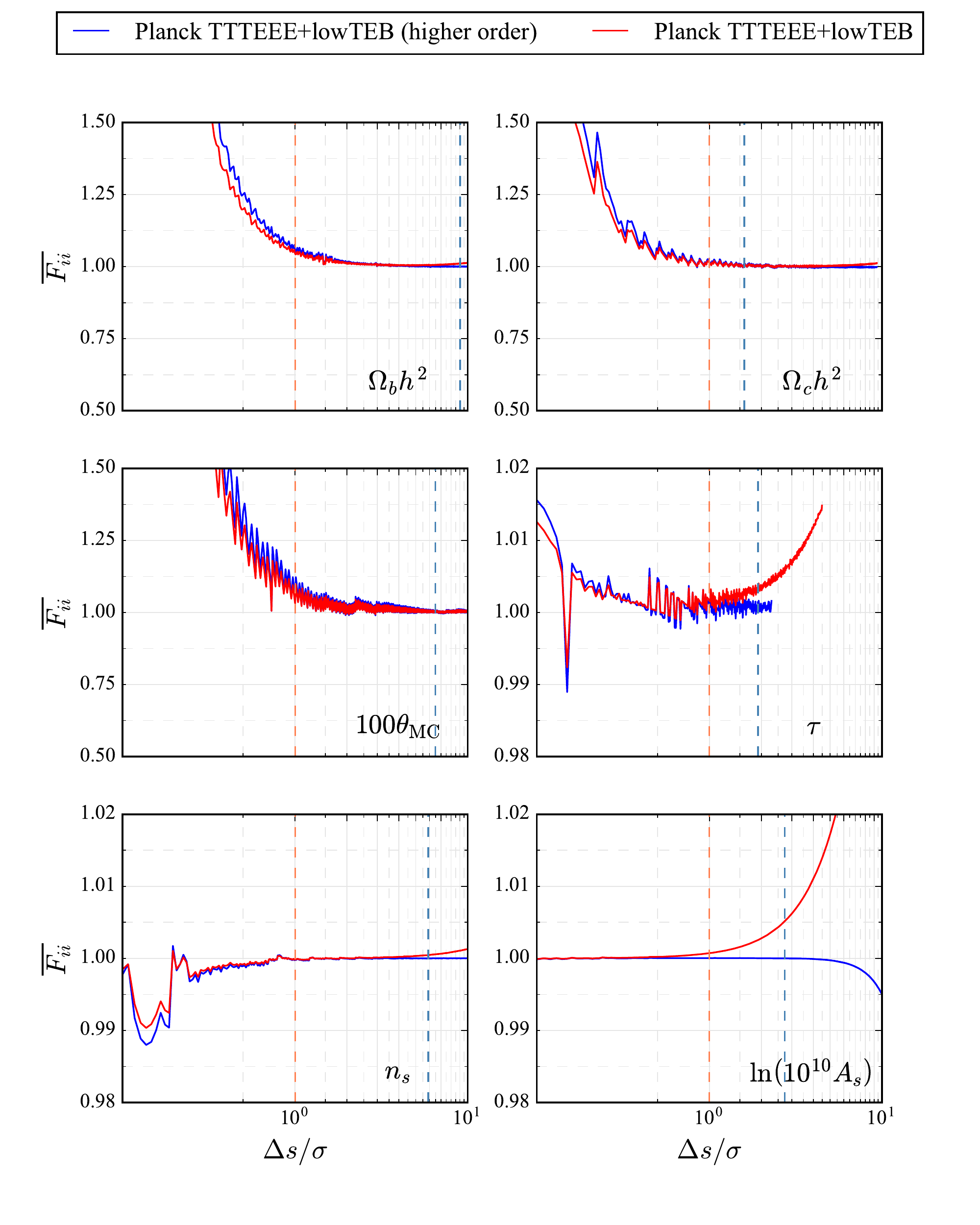}
\caption{Diagonal Fisher elements for the standard parameters using their fiducial best fit values from \planck as a function of step size for the derivatives. Note that the derivatives are in terms of $\sigma$. The orange dashed line represents the $1\sigma$ limit whilst the blue dashed line represents the step sizes we use in our analysis.}
\label{fig:stability_fisher}
\end{figure}

To decorrelate standard parameters, we need to make sure that the derivatives in the likelihood are stable before we proceed. The change in likelihood $\DelL$ values are shown for the standard \LCDM parameters in Fig.~\ref{fig:stability_curve}. The diagonal Fisher elements are shown for a range of step sizes, as a function of their fiducial \planck error, $\sigma$ in Fig.~\ref{fig:stability_fisher}. 

In Fig.~\ref{fig:stability_curve}, the parabolic nature of $\DelL$ is well-defined for most of the parameters except for $\omb$ and $\omc$, where there are some noisy elements. 
For Fig.~\ref{fig:stability_fisher}, the matter power spectrum parameters, $\ns$ and $\As$ were larged unaffected by the choice of step size. This is due to the smooth nature of their likelihood curves in Fig.~\ref{fig:stability_curve}. Note that the erratic behaviour of $\ns$ in Fig.~\ref{fig:stability_fisher} for very small $\Delta s/\sigma$ is most likely due to the slight deviation from the maximum of the likelihood where the derivatives are evaluated.
For $\As$, we need to select $\Delta s\simeq3\sigma$ so that we avoid the parabolic tail for larger values. Given the precise nature of $\thetaMC$, the choice was made to use a step for $\Delta s\simeq5\sigma$. 
The density parameters $\omb$ and $\omc$ display noisy behaviour below $\Delta s \simeq2\sigma$. The sweeping behaviour of the diagonal Fisher element can be attributed to the small scale noisiness on the log-likelihood curves in Fig.~\ref{fig:stability_curve}. This is something that we could not rectify due to the noisiness coming from the likelihood, however, this does not affect our final derivatives as we use a more stable step size. 
For $\tau$, as shown in Fig.~\ref{fig:stability_fisher}, the step size needed to be higher however the lower order stencil mixes the noise at small $\Delta s /\sigma$ with the non-linearities at high $\Delta s /\sigma$. This was remedied with the higher order stencil and we selected $\Delta s \simeq 2\sigma$. In this case, the noisy lower steps of the derivative mix with the non-linear parabolic tail as shown in the $\tau$ panel of Fig.~\ref{fig:stability_fisher}. This problem is alleviated in the higher order scheme (blue). The curves for $\tau$ in Fig.~\ref{fig:stability_fisher} end abruptly and this is due to the derivatives requring negative values of $\tau$ that are non physical (and also break the Boltzmann codes).  

\subsection{Marginalisation of the standard and nuisance parameters}
\label{app:marg}
In this section, we describe the marginalisation of the standard and nuisance parameters. The full set of 31 parameters we have marginalised over are shown in Table~\ref{tab:parameters}. The fiducial values of each of these parameters reflect the maximum likelihood estimators, using the {\tt BOBQYA} maximisation routine in {\tt CosmoMC}, for every given dataset (both for Fig.~\ref{fig:stability_curve} and Fig.~\ref{fig:stability_fisher}). 

We have to isolate the perturbations from the standard and nuisance parameters matrix $F_{ss}$. Using linear algebra rules, we can rewrite the Fisher matrix for the perturbations $F^{-1}_{pp}$ as,
\begin{equation}
  \left(F^{-1}\right)_{pp} = \left(F_{pp}-F_{ps}F_{ss}^{-1}F_{sp}\right)^{-1}.
  \label{eq:Fpp}
\end{equation}
Here we sidestep inversion issues arising from the size of the Fisher matrix. We also negate any numerical relics that are linked to the weakly constrained corners of the Fisher matrix $z_i \simeq \{300,2500\}$ creating a singular-like matrix using the method described in \eqref{eq:Fpp}.

As a result, we only need to invert a smaller, stable standard parameter Fisher matrix. The block inverted perturbation matrix $(F^{-1})_{pp}$ can then be diagonalised to generate marginalised eigenmodes. This works given that the eigenvectors for a given matrix are the same as its inverse. In principle, this decorrelates the changes in the free electron fraction due to the basis functions from the effects that standard parameters have on the ionization history \citep[see][]{Farhang2011}. 

\begin{table}
  \setlength{\tabcolsep}{5pt}
  \renewcommand{\arraystretch}{1.2}
  \centering
  \begin{tabular}{l | c }
    \hline\hline
    {\bf Parameter} & {\bf Symbol} \\
    \hline
Baryonic matter density & $\Omega_b h^2$ \\ 
Cold dark matter density & $\Omega_c h^2$ \\ 
Angular scale of last scattering surface & $100\theta_{MC}$ \\ 
Reionization optical depth & $\tau$ \\ 
Spectral amplitude of initial perturbations & ${\rm{ln}}(10^{10} \As)$ \\ 
Spectral index of initial perturbations & $n_s$ \\ \hline
Planck absolute calibration & $y_{\rm cal}$ \\ 
CIB contamination for 217GHz & $A^{CIB}_{217}$ \\ 
Cross correlation between the SZ and CIB signals & $\xi^{tSZ-CIB}$ \\ 
Contamination in the thermal SZ signal at 143 GHz & $A^{tSZ}_{143}$ \\ 
Point source contributions from frequency band  & $A^{PS}_{\mathcal{F}}$ \\
\qquad $\mathcal{F}\in\{100,143,217\}\,{\rm GHz}$ & \\
Contamination from the kinematic SZ signal & $A^{kSZ}$ \\ 
Dust contamination for a given $X\in \{TT,TE,EE\}$ & $A^{{\rm dust}X}_{\mathcal{F}}$ \\
\qquad and given  frequency band $\mathcal{F}\in \{100,143,217\}\,{\rm GHz}$ & \\
Relative calibration between 100 GHz and 143 GHz channels & $c_{100}$ \\ 
Relative calibration between 217 GHz and 143 GHz channels & $c_{217}$ \\ \hline\hline
  \end{tabular}
  \caption{The \planck 2015 parameters that are marginalised with the perturbation basis functions, describing the full Fisher matrix for the direct method. Note that the point source and dust contributions contain cross-correlations amongst the maps and the frequency bands as well.  \citep{Planck2015like}}
  \label{tab:parameters}
\end{table}

\subsection{Optimisation of the direct likelihood method}
\label{app:optimise}
One subtlety to point out here is the numerical ambiguity of the maximum likelihood estimation. For our likelihood method the accuracy of the results relies on the noise limit of the likelihood code as well as how the likelihood behaves when we are away from the maximum likelihood value. This has been discussed previously in the literature when MCMC simulations were not as readily available as maximum likelihood estimators \citep{Tegmark2004}. We tested the effective parameter distances from the maximum likelihood as shown in Fig.~\ref{fig:stability_curve}, however using linear algebra techniques that have been derived in the literature, we found the differences were negligible compared to the steps we are using in the Fisher analysis \citep[see][for detailed derivations]{Hobson2002}. Further studies on applying likelihood minima to Fisher matrix technqiues are discussed in the recent release of {\tt MontePython 3} \citep{Brinckmann2018}.

\section{Other likelihood contours for parameter extraction} 
\label{app:contours}
\begin{figure}
  \centering
  \includegraphics[width=.93\linewidth, trim={0 5 0 0},clip]{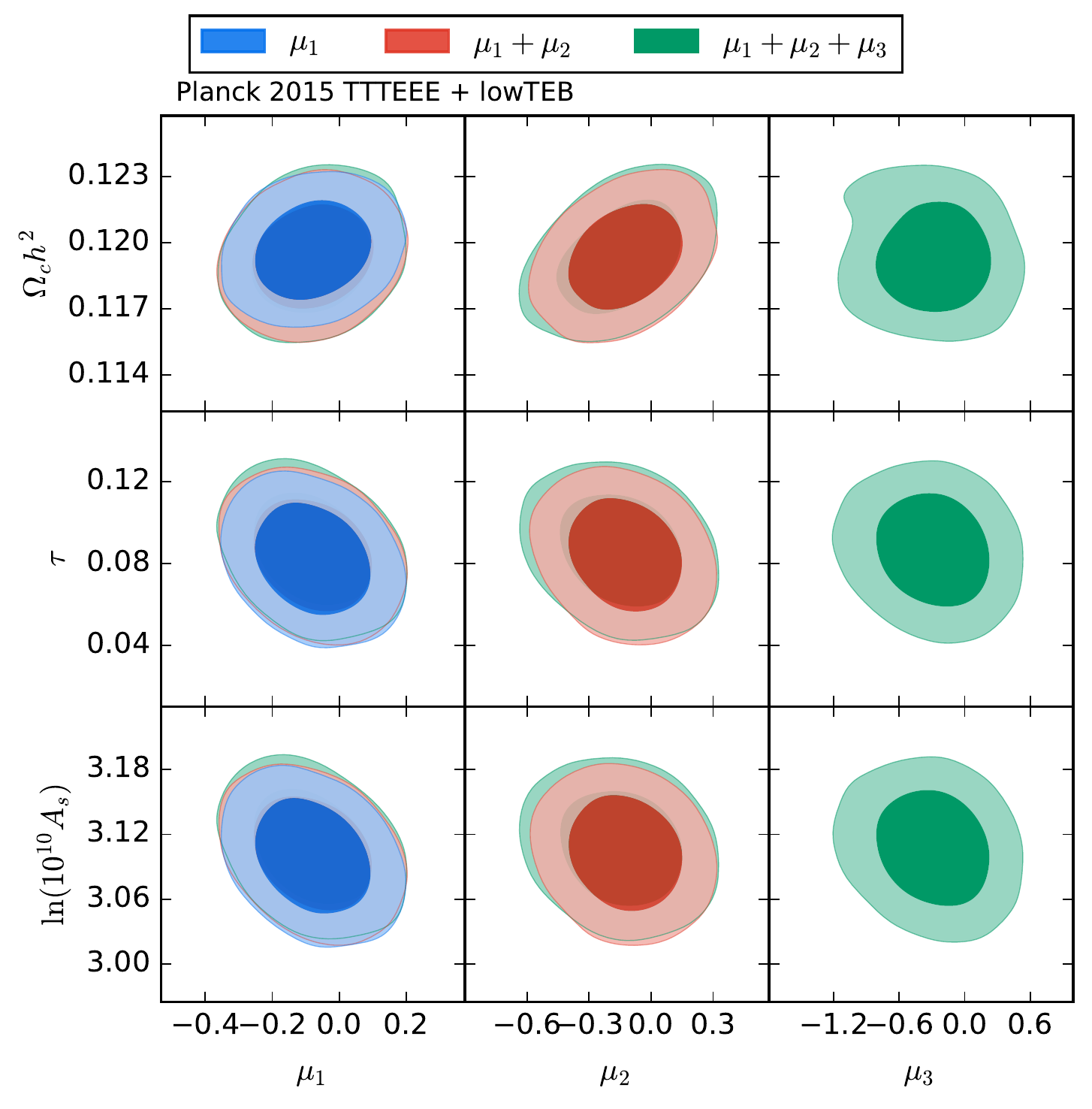}
  \caption{Posterior contours for the \planck TTTEEE + lowTEB eigenmodes. Here we show the contours for the first three recombination eigenmodes $\mu_{1,2,3}$ alongside the standard $\Lambda$CDM parameters $\left\{\omc,\tau,\As\right\}$ that were not highlighted in Fig.~\ref{fig:contours_degen}.}
  \label{fig:contours_extra_planck}
\end{figure}

To be completely transparent about parameter degeneracies, we have included the other correlations between the eigenmodes and standard parameters $\{\omc,\tau,\As\}$. In Fig.~\ref{fig:contours_extra_planck}, the posterior contours for these parameters are shown, where the degeneracies are noticeably uncorrelated. This is also shown in the marginalised limits in Table~\ref{tab:planck}. Though $\omc$ affects the recombination history, it has been almost fully decorrelated from the eigenmode parameters. The eigenmodes are decorrelated from $\tau$ and $\As$ due to the changes being focussed on the free electron fraction.

\label{lastpage}
\end{document}

%% file: Befehle.tex
\newcommand{\gsim}{\gtrsim}

\newcommand{\TCMB}{T_{\rm CMB}}

\newcommand{\xe}{x_{\rm e}}

\newcommand{\id}{{\,\rm d}}

\newcommand{\beq}{\begin{equation}}   %

\newcommand{\eeq}{\end{equation}}   %

\newcommand{\beqa}{\begin{eqnarray}}   %

\newcommand{\eeqa}{\end{eqnarray}}   %

\newcommand{\beal}{\begin{align}}
\newcommand{\enal}{\end{align}}

\newcommand{\bspl}{\begin{split}}

\newcommand{\espl}{\end{split}}

\newcommand{\bsub}{\begin{subequations}}

\newcommand{\esub}{\end{subequations}}

\newcommand{\bmulti}{\begin{multline}}   %

\newcommand{\beqm}{\begin{mathletters}}   %

\newcommand{\eeqm}{\end{mathletters}}   %

\newcommand{\me}{m_{\rm e}}

